\documentclass[11pt]{article}
\usepackage[a4paper,hmarginratio=1:1,vmarginratio=2:3,
totalwidth=15.5cm,totalheight=22.8cm]{geometry}
\usepackage{bm,epstopdf,epsfig,amsmath,amssymb,amsfonts,colordvi,latexsym,comment,cancel,
verbatim}
\usepackage{graphicx,graphics}
\usepackage[font=md,captionskip=8pt]{subfig}
\usepackage[usenames,dvipsnames]{color}

\usepackage{setspace}

\newcommand{\cL}{{\cal L}}

\newcommand{\cO}{{\cal O}}
\newcommand{\cP}{{\cal P}}

\newcommand{\cZ}{{\cal Z}}

\newcommand{\oD}{{\overline D}}

\newcommand{\ra}{\rightarrow}
\newcommand{\be}{\begin{equation}}
\newcommand{\ee}{\end{equation}}
\newcommand{\bea}{\begin{eqnarray}}
\newcommand{\eea}{\end{eqnarray}}
\newcommand{\Ra}{\Rightarrow}

\newcommand{\baa}{\begin{array}}
\newcommand{\eaa}{\end{array}}

\long\def\symbolfootnote[#1]#2{\begingroup
\def\thefootnote{\fnsymbol{footnote}}\footnote[#1]{#2}\endgroup}
\begin{document}

\thispagestyle{empty}
\begin{flushright}
CERN-PH-TH-2015-048 \\
CPHT-RR-008.0315\\
\end{flushright}

 \vspace{0.5 cm}

\begin{center}
{\Large {\bf Effective operators in SUSY, superfield constraints

\bigskip
  and searches for a UV completion}} 

 \medskip
 \vspace{1.cm}
\textbf{E. Dudas$^{\,a,b,}$\symbolfootnote[2]{E-mail: 
emilian.dudas@cpht.polytechnique.fr,dumitru.ghilencea@cern.ch},
\,\, D. M. Ghilencea$^{\,c,d\,}$}

\bigskip
$^a$ {\small Centre de Physique Th\'eorique,  \'Ecole Polytechnique, CNRS, 91128 Palaiseau, France}

$^b$ {\small Deutsches Elektronen-Synchrotron DESY, 22607 Hamburg, Germany}

$^c$ {\small CERN Theory Division, CH-1211 Geneva 23, Switzerland}

$^d$ {\small Theoretical Physics Department, National Institute of Physics and}

{\small  Nuclear Engineering (IFIN-HH) Bucharest, MG-6 077125,  Romania}
\end{center}

\begin{abstract}
\noindent
We discuss the role of a class  of higher dimensional
operators in 4D N=1 supersymmetric  effective theories. 
The  Lagrangian in such theories is an expansion in momenta 
below the scale of ``new physics'' ($\Lambda$) and contains the effective operators
generated by integrating out the ``heavy states'' above $\Lambda$
present in the UV   complete theory.
We go beyond the ``traditional'' leading order in this 
momentum expansion (in $\partial/\Lambda$).  
Keeping manifest supersymmetry and using  superfield {\it constraints} 
we show that the  corresponding higher dimensional (derivative) operators 
in the sectors of  chiral, linear and vector superfields of a Lagrangian can be
``unfolded'' into  second-order operators.
The ``unfolded'' formulation   has only polynomial interactions 
and additional massive  superfields,  some of which are ghost-like 
if the effective operators were {\it quadratic} in fields. 
Using this formulation, the UV theory emerges naturally and fixes
the (otherwise unknown) coefficient and sign of the initial (higher derivative)
operators.
Integrating the massive fields of the ``unfolded'' formulation
 generates an effective theory with only 
polynomial  effective interactions relevant for phenomenology. 
We also provide several examples of ``unfolding'' of  theories
with higher derivative {\it interactions} 
in the gauge or matter sectors  that are actually  ghost-free.
We then illustrate  how our method can be applied
even when including  {\it all orders} in the momentum expansion,
by using an infinite set of superfield constraints and an
iterative procedure,  with similar results.
\end{abstract}

\newpage

\tableofcontents{}

\section{Introduction}

Effective field theories are our main tool for studying physics at high scales,
like the  physics beyond the Standard Model. There are two reasons for this.
One reason is the absence  of a fundamental theory  (UV completion).
The second reason is that these theories are  convenient for
 practical purposes:  we do not need them if we have the 
full theory and are able to compute everything in such case; 
but effective theories make calculations much easier 
by  focusing on the  {\it relevant parameters} for the physics at 
the (momentum) scale investigated.  
Then shorter distance physics can  be ignored together with all particles too heavy to 
be produced at this scale. Eliminating ({\it i.e.} integrating out)
these particles simplifies the calculation. 
The result  is  a non-renormalizable theory (even if initial theory was
  renormalizable), in which all nontrivial effects of heavy particles 
appear in operators with dimensions $d>4$ \cite{wyler,georgi}. In the full theory, these effects 
are included in the non-local  interactions obtained by integrating out 
``heavy particles''.
But in the effective theories one replaces non-local interactions with virtual 
particles exchange by  a set of {\it local} interactions such as to give the same 
low energy physics.  The high energy behaviour is  affected, so the effective 
theory is only valid at momenta below the mass of the ``heavy particles''. This mass 
is the effective theory cut-off ($\Lambda$).

 The Lagrangian of the effective theory then
contains just  local interactions,  obtained from an expansion in momenta below 
this scale,  i.e. in powers of $\partial/\Lambda$,  up to some  finite order. 
Keeping {\it all orders} in momenta  leads to  a {\it non-local}
 theory equivalent to the 
original full theory\footnote{There is nothing wrong in using this non-local theory 
for calculations \cite{georgi}.
For  the differences between  this non-local and 
effective local theories, see \cite{georgi}; we shall meet 
such an example in Section~\ref{s7} and the  Appendix.}.
The effective Lagrangian is analytic in $\partial/\Lambda$ in the region relevant
to the low energy theory and it can be dealt with in any  finite order 
in the momentum expansion. This is  a local Lagrangian that one is using.
This picture is so familiar that it is usually implicitly assumed,
but we reminded  it to make clear our set up. For a review see \cite{georgi}.

In this work we would like to investigate   such effective 
theories beyond the first leading term of the momentum expansion.  
This is relevant when the momentum is closer to $\Lambda$.
Using this picture we also try to infer a UV theory at 
scales   above the ``heavy particles'' mass. By  
``UV theory'' we mean a two-derivative ghost-free theory with  only 
polynomial interactions in superfields. If this theory is  renormalizable
we refer to it as a UV completion\footnote{An example of a UV theory 
is 4D N=1 supergravity (UV incomplete).}.

Let us formulate the above picture and our goals in a more precise way. 
In 4D SUSY effective theories the study is often strongly
restricted to  Kahler potentials $K$ and superpotentials $W$ 
that depend only on the superfields $\Phi^i$. 
But one often encounters  cases when  $K$ and $W$  depend on more 
general arguments such as
the superderivatives $D_\alpha$ acting on the superfields
i.e. $K\!=\!K(\Phi^i, \Phi_j^\dagger, \overline D\Phi^\dagger_k, D\Phi^l, \overline D^2 
\Phi^\dagger_m, D^2\Phi^n,...)$ and $W\!=\! W(\Phi^i, \overline D^2  \Phi^\dagger_i,...)$.
In component fields, this action contains powers  of  
$\partial^\mu$, $\Box$, etc. These  account for the momentum 
expansion $\partial/\Lambda$ giving  the  effective theory mentioned. 
$\Lambda$ is here related to Kahler curvature.

For the two-point Green functions (propagators)
the presence of an expansion in powers of $\partial/\Lambda$ 
(giving higher dimensional derivative operators) can 
lead to additional poles; some of these are ghosts,
(super)fields of negative kinetic terms.  
There is nothing   ``pathological'' about their presence here.
They are just artefacts of the
effective approach that are eliminated by the field equations
or non-linear field redefinitions, to leave a ghost-free effective theory.
As mentioned, the corresponding (derivative) operators are a
common presence in the low energy limit of  the 
 UV  theory, after integrating out the ``heavy particles''.
They are thus  related to the UV completion  of the effective  action.
Such operators are also  present in the interactions terms. 
When  scalar fields in these interactions
develop vev's,  these terms can  contribute  
to the two-point Green functions, with  similar effects (additional 
poles and ghosts). These operators are studied below.

Let us detail how such operators emerge
when  classically integrating out  massive states.
Consider a simple  (UV complete) Lagrangian
\medskip
\bea
\cL=\int d^4\theta\,\,
\Big\{ \Phi_0^\dagger\,\Phi_0 +\chi^\dagger \chi\Big\}
+
\Big\{\int d^2\theta \Big[ (1/8)\,\Lambda \chi^2 + (1/4)\,m_0\,\Phi_0 \chi +W(\Phi_0)\Big]+h.c.
\Big\}
\label{i1}
\eea

\medskip\noindent
For large $\Lambda$, the superfield $\chi$ has a large mass, so it can be 
integrated out using  its equation of motion $-\overline D^2\chi^\dagger +\Lambda\chi + m\Phi_0=0$. 
This has an iterative solution
\medskip
\bea\label{i2}
\chi=\frac{-m_0}{\Lambda}\Phi_0 +\frac{-m_0}{\Lambda^2}\,\overline D^2\Phi_0^\dagger 
+\frac{-m_0}{\Lambda^3} \overline D^2 \,D^2 \Phi_0+
\cdots
\eea

\medskip\noindent
Thus $\chi$ is an infinite series in $(\partial/\Lambda)$.
This solution is used  back in $\cL$ to give 
\medskip
\begin{equation}\label{i3}
\cL_{\rm eff}\!=\! \!\int\! 
 d^4\theta \,\Phi_0^\dagger\Big[
1-\frac{16 m_0^2}{\Lambda^4}\,\Box\Big]\Phi_0
+ \Big\{\!\!
\int\! d^2\theta\,\Big[W(Z^{\frac{1}{2}}\,\Phi_0)
+\frac{2 m^2_0}{\Lambda^3}\,\Phi_0 \Box\Phi_0\Big]\!+{\rm h.c}\!\Big\}
+\cdots\end{equation}

\medskip\noindent
where we replaced   $\overline D^2 D^2\ra -16 \Box$ and $Z=1/(1+m^2/\Lambda^2)$.
So a simple decoupling of a
massive state generated  the  $\Box$-operators. Actually,
due to  eq.(\ref{i2}), eq.(\ref{i3}) contains an {\it infinite} series 
in momentum expansion ($\partial/\Lambda$), from the initial
 renormalizable theory.
One usually truncates this series to a low(est) order, as in (\ref{i3}).
Then $\Box$-operators are often eliminated 
by simply  using the leading order equations of motion\footnote{In SUSY 
theories with higher derivative terms, the auxiliary fields  
become dynamical in most cases.}
(for the non-SUSY case, see \cite{hawking,hawking2}).

Such operators are also  generated dynamically by  compactification, as loop counterterms. 
They can be  generated by bulk (gauge) interactions to give
the F-term below  that contributes  to a  one-loop running 
of the 4D  effective gauge coupling in orbifolds 
\cite{Oliver:2003cy,Ghilencea:2006qm,Ghilencea:2003xj,Ghilencea:2006ye,GrootNibbelink:2005vi,
GrootNibbelink:2006ad,Nibbelink:2006eq} 
\bea \label{1q}
\delta\cL\supset
R^2\int d^4\theta \,\,\Phi^\dagger \Box\Phi
+
\Big\{R^2\int d^2\theta \,\,W^\alpha\Box W_\alpha+h.c.\Big\}
\eea

\medskip\noindent
where $1/R$ is the compactification scale and $W^\alpha$ is the gauge field strength.
Regarding the D-term, it is generated by  superpotential (Yukawa) interactions 
localised at the 4D fixed points  of an orbifold
(it can be a Higgs mass counterterm in such orbifolds
 \cite{Ghilencea:2004sq,Ghilencea:2005hm}).
So models with extra dimensions contain 
 such effective operators\footnote{
The relation of these results to  string theory  
is discussed in \cite{Ghilencea:2006qm,GrootNibbelink:2006ad,Nibbelink:2006eq}.}
at one-loop. Such operators can also be
 present in other compactifications  (Randal-Sundrum, etc). 
In  conclusion, these operators are   common  and are related 
to the  UV regime.

The main   goal of this work is to clarify two problems
for general effective theories:

\noindent
{\bf 1):} to obtain a better understanding of the higher order terms 
in momentum expansion,  and  then  ``remove'' these 
higher derivative operators from the effective Lagrangian.
To do this   we show that one can reformulate (``unfold'') such a Lagrangian
into  a {\it second-order}  Lagrangian i.e. without  higher dimensional (derivative)
operators. This is what we mean by ``unfolding''.
This result is interesting since 
a two-derivative formulation of a theory is easier to handle 
and the ``unfolded'' formulation is  a first step towards
 identifying a (ghost-free)  UV theory.

\noindent
{\bf 2):}  to identify  a two-derivative ghost-free  UV theory
of the effective theory with such operators.
It is of strong interest to find a UV quantum consistent theory leading
in the infrared to these (effective) operators and to fix
in this way  their (otherwise unknown)  coefficient and sign,
in agreement with constraints derived from analyticity 
and causality \cite{nima}. 
The ``unfolded'' formulation will help us to  achieve this.
For a related discussion see
\cite{Antoniadis:2007xc,Antoniadis:2008ur,Antoniadis:2008es,Antoniadis:2009rn}.

We show how to  ``remove''  the $\Box$-operators 
acting on  \,chiral,  vector or  linear superfields,
in a consistent way, while preserving manifest supersymmetry 
(in a superfield language). {We show how (effective) 
theories that contain such operators can be ``unfolded'' into second order theories
with only polynomial interactions 
and with additional {\it massive} states, sometimes with negative kinetic 
terms (ghosts)}. For this, one eliminates these operators by suitable 
superfield {\it constraints}. For example, in the chiral sector
such  constraints replace each  $\overline D^2\Phi^\dagger$ by a superfield 
$m\,\Phi^\prime= \overline D^2\Phi^\dagger$. Here $m$ is a small {\it arbitrary}
 scale of the theory that enforces the constraint.
All superderivatives are thus eliminated, to find a second-order theory.
The method can be iterated to higher orders in $D_\alpha, \oD^{\dot\alpha}$
and can also be applied to non-derivative effective operators.
Subsequent integration of these massive states leads to a 
theory with effective {\it polynomial} operators only and
this formulation  can be used for phenomenology.

We then show how  the ``unfolded'' formulation helps us identify a UV theory
of the initial effective theory. 
In the above models the initial effective operator was quadratic in fields. 
We also study other models with higher derivative interactions
 that are {\it ghost-free}:
 a chiral superfield model with such operators, a model of  Dirac gaugino 
masses and  a supersymmetric version of the Euler-Heisenberg Lagrangian. 
We  ``unfold'' them and then find their UV theory.

The plan of the paper is as follows.
For the operators of eqs.(\ref{i3}), (\ref{1q}) acting on 
chiral, linear and gauge superfields 
the ``unfolding'' method is done in Sections~\ref{s2}, \ref{linear} and \ref{s4}.
 This extends our study in \cite{Antoniadis:2007xc}.
For related discussions see \cite{K1,K2,K3,K4,BP,K5,K6,K7,K8,K9,K10,K11}.
In each of these Sections we  use the ``unfolded'' formulation to identify a 
UV theory (or UV completion) that generates
at low energy the effective operators considered. 
Section~\ref{s2}  also contains the unfolding
of a higher-derivative ghost-free chiral superfield model. Section~\ref{s4}
 contains ghost-free examples leading to Dirac gaugino masses and to the
Euler-Heisenberg gauge Lagrangian.  
Section~\ref{s7} comments on  how to treat more general cases.
The Appendix presents the  ``unfolding'' of  a theory with an (known) infinite series
of superderivatives. Using {\it an infinite} set of superfield constraints it is shown that
even  in this case there exists an  ``unfolded'' version
with only  polynomial (d=4) interactions 
and an infinite set of extra (massive) superfields. 
Truncating the ``unfolded'' theory to a  number  of such fields is equivalent 
to the truncation to a corresponding power in $(\partial/\Lambda)$ of
 the momentum expansion.

\section{Effective operators in the chiral  sector}
\label{s2}

\subsection{``Unfolding'' the  effective operators}

Let us first consider the case the $\Box$-operators, eqs.(\ref{i3}), 
(\ref{1q})  in the matter sector \cite{Antoniadis:2007xc}.
Consider
\medskip
\bea\label{e0}
\cL=\int d^4\theta\, \Big[\,\Phi^\dagger_1 \Phi_1 
+\frac{\rho}{\Lambda^2} \Phi^\dagger_1 \Box\Phi_1\Big]
+\Big\{\int d^2\theta \,
\Big[\,\,\frac{\sigma}{\Lambda}\Phi_1\Box\Phi_1+W(\Phi_1)\Big]+{\rm h.c.}\Big\} 
+\cO(1/\Lambda^3)\eea

\medskip\noindent
where $\rho, \sigma=\cO(1)$ are independent.
We replace $\Box$ by $(-1/16)\overline D^2 D^2$.
Further, introduce $\Phi_2$
\bea\label{e2}
\overline D^2\Phi_1^\dagger - m\, \Phi_2=0.
\eea
%
where $m$ is a real,  small but  arbitrary mass scale of the theory.
This is a superfield constraint that we add to $\cL$,
using a Lagrange multiplier superfield\footnote{This Lagrange multiplier method is similar 
to the one used in \cite{buchmuller} in a case without higher derivatives.} $\Phi_3$.
Then
\bea\label{e1}
\cL&=&\int d^4\theta \,\Big[\Phi_1^\dagger \Phi_1
 -\frac{\rho\,m^2}{16 \Lambda^2} \,\Phi_2^\dagger \Phi_2\Big]
\nonumber\\[4pt]
&+&
\Big\{\int d^2\theta \,\Big[W(\Phi_1)-
\frac{\sigma\,m}{16\Lambda}\, \Phi_1 \overline D^2 \Phi_2^\dagger
- \frac{m}{4\,\Lambda}\, \Phi_3\, (\overline D^2\Phi_1^\dagger - m\Phi_2)
\Big] + {\rm h.c.}\Big\} +\cO(1/\Lambda^3)
\eea

\medskip\noindent
The field equation for $\Phi_3$  recovers the constraint.
Also, this constraint is implemented with a coefficient $1/\Lambda$ 
because it must be removed when $\Lambda\ra \infty$, while
$m$ is an unphysical parameter that 
 restores the mass dimension (at the end of the calculation we 
take $m\ra 0$). $\cL$ becomes
\medskip
\bea\label{e3}
\cL&=&\int d^4\theta 
\,\Big[
\Phi_1^\dagger \Phi_1
-\frac{\rho\,m^2}{16 \Lambda^2} \,\Phi_2^\dagger \Phi_2
+\frac{m}{4\Lambda} \,(\sigma\,\Phi_1^\dagger\Phi_2+ \sigma^* \Phi_1\Phi_2^\dagger)
+\frac{m}{\Lambda}\,(\Phi_1^\dagger \Phi_3+\Phi_1\Phi_3^\dagger)
\,\Big]
\nonumber\\[4pt]
&+&
\Big\{\int d^2\theta\, \Big[W(\Phi_1)+ \frac{m^2}{4 \Lambda}\, \Phi_2\,\Phi_3
\Big] + {\rm h.c.}\Big\} +\cO(1/\Lambda^3)
\eea

\medskip\noindent
 The (hermitian) matrix $k_{ij}$ of the kinetic (D-)terms 
$\Phi_i^\dagger\,k_{ij}\Phi_j$ has $\det k_{ij}=\rho\, m^4/(16\Lambda^4)$. 
Its real eigenvalues\footnote{The matrix is $k_{11}=1$, 
$k_{12}=\sigma\, m/(4\Lambda)=k_{21}^*$, 
$k_{13}=m/\Lambda=k_{31}^*$, 
$k_{22}=-\rho\, m^2/(16 \Lambda)$, 
$k_{23}=0=k_{32}=k_{33}$.}
control the  nature of the superfields $\Phi_{1,2,3}$: positive (negative) 
eigenvalues correspond to particle-like (ghost-like)
superfields, respectively. We therefore have:

\medskip\noindent
a) If $\rho=1$ then we have two negative and one positive eigenvalue, the latter corresponding
to the original particle-like degree of freedom. Two superghosts are present, one due to the
operator $\Box$, the second because auxiliary $F_1$ of $\Phi_1$
 became dynamical, thus an extra d.o.f. is present
which, by supersymmetry, demands the presence of an extra (ghost) superfield.

\noindent
b) If $\rho=-1$, one has one negative (1 ghost) and two positive (2 particles) 
eigenvalues\footnote{We cannot have 3 negative eigenvalues, since we 
had one positive value to begin with (the initial particle)}. 

\noindent
c) If $\rho=0$ then one eigenvalue is 0, one is positive and one is negative.
Thus we have one ghost and one particle superfields.
All these eigenvalues are the roots of
\bea
-\nu^3+\big[1-\rho\,m^2/(16\Lambda^2)\big]\,\nu^2 
+(1+\rho/16+\vert \sigma\vert^2/16)\,\nu \, m^2/\Lambda^2
{+}
\rho\, m^4/(16\Lambda^4)=0
\eea

\medskip\noindent
The exact expressions can be obtained.
According to our discussion we have
 \bea
\nu_1>0, \qquad \nu_2<0, \qquad \nu_3\sim -\rho
\eea
The notation $\nu_3\sim -\rho$  means 
$\nu_3$ has the sign of $(-\rho)$ and is 0 if $\rho=0$. 
This covers all cases discussed above.
For a complete analysis, we also bring $\cL$  to canonical form using a transformation
to  $\Phi^\prime_i=u_{ij}\,\Phi_j$ with 
 diag$(\nu_1,\nu_2,\nu_3)=u\, k\, u^\dagger$ and $u_{ij}$ 
unitary. With the notation  $z_{kj}= u_{k2}^* u_{j3}^*\,m^2/(4\Lambda)$ and
after rescaling  $\Phi^\prime_i=\tilde\Phi_i/\sqrt{\vert \nu_i\vert}$, 
($\nu_i\not=0$), one finds
\medskip
\bea\label{e4}
\cL\!=\!
\int\! d^4\theta \Big[
\tilde\Phi_1^\dagger \tilde \Phi_1
- \tilde \Phi_2^\dagger \tilde \Phi_2
- \rho\,\tilde \Phi_3^\dagger \tilde \Phi_3\Big]
\!+\!
\Big\{\!\int d^2\theta
\Big[ W\Big(\frac{u_{j1}^*\tilde\Phi_j}{\sqrt{\vert \nu_j\vert}}\Big)
\!+ \!\frac{z_{ij} \tilde \Phi_i\tilde\Phi_j}{\sqrt{\vert\nu_i\vert \,\vert\nu_j\vert}} \Big]
\!+\!{\rm h.c.}\Big\}
\!+\!\cO(1/\Lambda^3)
\eea
 
\medskip\noindent
The initial $\cL$ was ``unfolded'' into a second order theory, with extra
superfields $\tilde\Phi_{2,3}$ 
(of mass $\sim\Lambda$, see later) and at least one of them having ``wrong''-sign 
kinetic term (superghost). None of the auxiliary fields is  dynamical anymore. 
The effective operators are not present anymore and all interactions are 
polynomial, up to  $\cO(1/\Lambda^3)$.

If $\sigma=0$ then, with $\rho=\pm1$: 
\medskip
\bea\label{e5}
\cL\!=\!
\int d^4\theta \Big[
\tilde\Phi_1^\dagger \tilde \Phi_1
- \,\tilde \Phi_2^\dagger \tilde \Phi_2
-\rho\, \tilde \Phi_3^\dagger \tilde \Phi_3\Big]
\!+\!
\Big\{\!\int d^2\theta
\Big[ W\big(\tilde \Phi_1-\tilde \Phi_2\big)
+ \Lambda \,\tilde \Phi_2\tilde\Phi_3 \Big]
\!+\!{\rm h.c.}\Big\}
\!+\!\cO(1/\Lambda^3)
\eea
 
\medskip
If instead $\rho=0$ and $\sigma=\pm 1$ 
\medskip
\bea\label{e6}
\cL\!=\!
\int d^4\theta \Big[
\tilde\Phi_1^\dagger \tilde \Phi_1
- \tilde \Phi_2^\dagger \tilde \Phi_2\Big]
+ \Big\{\!\int d^2\theta
\Big[ W\big(\tilde \Phi_1-\tilde \Phi_2\big)
+ (1/4)\,\sigma\,\Lambda\, \tilde \Phi_2^2\,
\Big]
\!+\!{\rm h.c.}\Big\}
\!+\!\cO(1/\Lambda^3)
\eea
 
\medskip\noindent
Similar expressions exist for the more general case when $\rho\not=0$ and $\sigma\not=0$ 
simultaneously. Note that in the arguments of $W$ in the last two equations
 and also inside the
square bracket of the F-terms there are additional terms $\cO(m/\Lambda)$ that 
we did not write since we  now set $m\ra 0$ because we do not have 
a constraint anymore. As a side remark, notice that 
the scalar potential is $V\!=\!\vert \tilde F_1\vert^2\!-\vert \tilde F_2\vert^2-
\rho\vert\tilde F_3\vert^2$ in eq.(\ref{e5}) and 
$V\!=\!\vert \tilde F_1\vert^2\!-\vert \tilde F_2\vert^2$
in eq.(\ref{e6}) where $\tilde F_i$ are the auxiliaries of superfields $\tilde\Phi_i$.
This allows  $V=0$ with broken {\it global} SUSY.

The effect of the original higher dimensional operators was then to introduce ghost superfields,
 of large  mass (of order  $\Lambda$) as shown by the F-terms in the last
 two equations\footnote{
The difference in the number of superghosts is because if $\rho\not=0$ the
auxiliary field of $\Phi$ becomes dynamical (unlike the case of $\rho=0$) and
by SUSY this brings an extra superfield in the ``unfolded'' Lagrangian.}.
Using the equations of motion one can integrate 
out the massive ghost superfields. 
For example in eq.(\ref{e6}), one uses the equation of motion for $\Phi_2$ 
\medskip
\bea
(-1/4)\,\overline D^2 \tilde\Phi_2-W^\prime(\tilde\Phi_1-\tilde\Phi_2)
+ (1/2)\,\sigma\Lambda\, \tilde \Phi_2=\cO(1/\Lambda^3)
\eea

\medskip\noindent
where the derivative of $W$ is wrt its shown argument.
This gives 
\medskip
\bea
\tilde\Phi_2=\frac{2\sigma}{\Lambda} \,W^\prime(\tilde\Phi_1-\tilde\Phi_2)
-\frac{1}{\Lambda^2}\overline D^2 W^\prime(\tilde\Phi_1-\tilde\Phi_2)
+\cO(1/\Lambda^3)
\eea

\medskip\noindent
One then expands  $W^\prime(\tilde\Phi_1-\tilde\Phi_2)=W^\prime
-\tilde\Phi_2\,W^{\prime\prime}+(1/2)\,\tilde\Phi_2^2 \,W^{\prime\prime\prime}
+\cO(1/\Lambda^3)$ where we introduced 
the notation $W^\prime\equiv W^\prime(\tilde\Phi_1)$ 
 $W^{\prime\prime}\equiv W^{\prime\prime}(\tilde\Phi_1)$. Therefore
\medskip
\bea
\tilde\Phi_2=\frac{2 \,\sigma}{\Lambda} 
 W^\prime
- \frac{1}{\Lambda^2}\,\overline D^2 W^{\prime \dagger}
- \frac{4}{\Lambda^2}\,W^\prime\,W^{\prime\prime}+
\cO(1/\Lambda^3)
\eea

\medskip\noindent 
Using this and expanding  $W(\tilde\Phi_1-\tilde\Phi_2)$ about 
$\tilde\Phi_1$ in eq.(\ref{e6}), one obtains $\cL_{\rm eff}$
\medskip
\bea\label{eer}
\cL_{\rm eff}=\!\!\int d^4\theta\,\, \Big[
\tilde \Phi_1^\dagger \,\tilde \Phi_1
-\frac{4}{\Lambda^2}\,\vert W^\prime\vert^2\Big]
\!+\!
\Big\{\!\!\int d^2\theta \Big[W
-\frac{\sigma}{\Lambda} W^{\prime \,2}+\frac{2}{\Lambda^2} W^{\prime\, 2}\,W''
\Big]\!+\!{\rm h.c.}\Big\}\!+\!\cO(1/\Lambda^3)
\eea

\medskip\noindent
where  $W\equiv W(\tilde\Phi_1)$ and $\sigma=\pm 1$. 
Thus effective operators re-emerge but are now
 {\it polynomial} in fields.
 $\cL_{\rm eff}$ is classically
 equivalent to starting $\cL$ in eq.(\ref{e0}) for $\rho=0$.
Even if $W^\prime$ may contain a linear dependence on $\tilde\Phi_1$, no ghost
can be generated in the D-term due to the suppression  $1/\Lambda^2$ 
(relative to dominant $\tilde\Phi_1^\dagger\tilde\Phi_1$). Eq.(\ref{eer}) can now
be used for phenomenology. 

For eq.(\ref{e5}) integrating
the superghosts $\tilde\Phi_2, \tilde\Phi_3$ of mass $\sim \Lambda$ 
is done similarly to find
\medskip
\bea\label{eer2}
\cL_{\rm eff}=\!\!\int d^4\theta\,\, \Big[
\tilde \Phi_1^\dagger \,\tilde \Phi_1
-\frac{\rho}{\Lambda^2}\,\vert W^\prime(\tilde\Phi_1)\vert^2\Big]
+
\Big\{\int d^2\theta\,\, W(\tilde\Phi_1)+{\rm h.c.}\Big\}+
\!\cO(1/\Lambda^3)
\eea

\medskip\noindent
which is equivalent to eq.(\ref{e0}) for $\sigma=0$. This is the formulation
that can be used for phenomenology.
In both examples there are no ghost superfields 
as asymptotic (final) states in the approximation  $\cO(1/\Lambda^3)$.
A similar result is obtained if both $\rho,\sigma\not=0$.

The method can be extended to more general cases and higher orders, etc.
An alternative to our approach that leads to results identical to those  in
eqs.(\ref{eer}), (\ref{eer2}), is to use in eq.(\ref{e0}) non-linear 
field  redefinitions to ``remove'' 
the derivative operators  \cite{Antoniadis:2009rn}.

\subsection{A ghost-free UV theory in the matter sector}
\label{s3}

The above result indicates  how a UV theory of 
the starting Lagrangian eq.(\ref{e0}) can be realised, that 
is ghost free and, in this case, also renormalizable (UV complete). 
If we ignore the form of the dimensionless constants and $\cO(1/\Lambda^3)$ terms,
  Lagrangian (\ref{e3})  contains only dimension-four operators. The only 
problem is that it has ghosts, so it is not UV complete. This is seen after its kinetic
terms are diagonalised, leading to  results (\ref{e4}) to (\ref{e6})
 in which ghost superfields emerge. 
Their presence is induced by the kinetic mixing in (\ref{e3}) that is dominant
in the D-term, due to the absence of diagonal kinetic terms for $\Phi_2$ and 
$\Phi_3$.  This indicates that a ghost free theory
should thus ``UV complete'' Lagrangian (\ref{e3}) by the addition 
of diagonal kinetic terms
\bea
\delta \cL=\zeta \, \Phi_2^\dagger\Phi_2+\eta\, \Phi_3^\dagger\Phi_3
\eea
%
with suitable values for real $\zeta$, $\eta$.
Let us examine the impact of these terms. The new $\cL$  becomes 
\medskip
\bea\label{w0}
\!\!\cL\!\!&=&\!\!\!\!\int\! d^4\theta \,
\Big[
\Phi_1^\dagger \Phi_1\!
+\Big(\zeta-\frac{\rho\,\xi^2}{16}\Big) 
\,\Phi_2^\dagger \Phi_2
+\!
\frac{\xi}{4}\,(\sigma\,\Phi_1^\dagger\Phi_2+ 
\sigma^* \Phi_1\Phi_2^\dagger)
\!+\!
\xi\,(\Phi_1^\dagger \Phi_3\!+\Phi_1\Phi_3^\dagger)\,
\!+\!
\eta\,\Phi_3^\dagger\Phi_3\Big]
\nonumber\\[4pt]
&+&
\Big\{\int d^2\theta\, \Big[W(\Phi_1)+ \frac{1}{4}\,m\,\xi\, \Phi_2\,\Phi_3\Big]
 + {\rm h.c.}\Big\} 
\eea

\medskip\noindent
with $\xi$ real\footnote{
$\xi$  is just a   dimensionless parameter ($\xi\ra m/\Lambda$ in Section~\ref{s2}).
In the  UV theory $\xi$ ($m$) becomes  physical.}.
Let us first show that this  UV completed  theory,
recovers at low energy the  Lagrangian in eq.(\ref{e0}).
The equations of motion for $\Phi_{2,3}$ are
\medskip
\bea
\Phi_3&=&
\frac{\sigma}{4\,m}\overline D^2\Phi_1^\dagger +\frac{1}{m^2\xi}\,
\Big(\zeta-\frac{\rho\,\xi^2}{16}\Big)
\overline D^2 D^2\Phi_1+\cO(1/m^3)
\nonumber\\[5pt]
\Phi_2&=&\frac{1}{m}\overline D^2\Phi_1^\dagger +\frac{\eta\,\sigma^*}{4\,m^2\xi}
\overline D^2 D^2\Phi_1+\cO(1/m^3)
\eea

\medskip\noindent
Using this back in $\cL$ gives, up to $\cO(1/m^3)$ terms
\medskip
\bea
\cL\!= \!\int\! d^4\theta \Big[
\Phi_1^\dagger\Phi_1\!+\!
\frac{16}{m^2}\,\Big(\frac{\rho\,\xi^2}{16}\!-\!\zeta\!-\!
\frac{\eta\,\vert\sigma\vert^2}{16}\Big)
\,\Phi_1^\dagger \Box\Phi_1\Big]
\!+\!
\Big\{\int d^2\theta
\Big[ W(\Phi_1)+\sigma^*\xi\,\Phi_1\Box\Phi_1\Big]
\!+\!{\rm h.c.}\!\Big\}
\eea

\medskip\noindent
This $\cL$ is identical to the original Lagrangian of eq.(\ref{e0}) provided that
\bea\label{w1}
\zeta=-\eta\,\vert\sigma\vert^2/16 
\eea
We thus  need only one extra parameter $\eta$ to find a  UV theory. 

As a UV theory, this theory must be ghost-free, so let us check under
what conditions this is true. 
This happens if we have positive values for the eigenvalues $\tilde \nu_{1,2,3}$ 
of the matrix of the quadratic form of the D-term in 
eqs.(\ref{w0}), (\ref{w1}); $\tilde\nu_{1,2,3}$ are the roots  of
\medskip
\bea
\label{roots}
-\tilde\nu^3\!&+&\!\tilde\nu^2\,\,(\eta+\zeta+1-\rho\,\xi^2/16)
+\tilde\nu\,\,\Big[\xi^2 (1+(\rho+\vert\sigma\vert^2+\eta\rho)/16)-\zeta-\eta-\eta\,\zeta\Big]
\nonumber\\
&+&
\!\!(\zeta-\rho\xi^2/16)\,\,(\eta-\xi^2)-\eta\,\xi^2\,\vert\sigma\vert^2/16=0
\eea

\medskip\noindent
with $\zeta$ as in eq.(\ref{w1}). Let us determine the 
values of the new coefficient  $\eta$ so that  $\tilde \nu_{1,2,3}\geq 0$.

\noindent
{\bf a).} if $\sigma=0$ and $\rho<0$ all roots $\tilde \nu_{1,2,3}\geq 0$
 provided that $\eta\geq \xi^2$. 

\noindent
{\bf b).} if $\sigma=0$ and $16/\xi^2>\rho>0$, all $\tilde \nu_{1,2,3}\geq 0$ if
 $\xi^2\geq \eta\geq (1+ \rho/16)\xi^2/(1-\rho\xi^2/16)$. 

\noindent
{\bf c).} if $\sigma=0$, for $\rho\geq 16/\xi^2$ there is no solution for $\eta$ for which 
all $\tilde \nu_{1,2,3}>0$. 

\noindent
{\bf d).} if $\sigma=\pm 1$ all roots are positive if one respects simultaneously the
following conditions:
$\eta\geq 16 \,(\rho\xi^2/16-1)/15$ and
$-\eta^2+\eta \,(15-\rho\xi^2)-\xi^2\,(17+\rho)\geq 0$ and
$-\eta^2-\eta \,\rho\,\xi^2+\rho\xi^4\geq 0$. 
In this last case, the solution for $\eta$ 
exists and depends on the exact values of  $\rho$ and $\xi$. 

For example, if
$\rho=1$ and $\vert\xi\vert\ll 1$, $-0.38 \xi^2\leq \eta\leq 0.62 \xi^2$. 
Thus, if both effective $\Box$ operators are present, a UV theory, free of ghosts exists
(all roots are positive).
For $\rho=0$, i.e. no $\Phi_1^\dagger\Box\Phi_1$ in original $\cL$,
there always exists a negative root.
As a result, if an $\cL$ contains only the  effective operator
$\int d^2\theta\Phi_1\Box\Phi_1$
there is  {\it no  ghost-free,  UV theory.}

From the above cases we select those when all roots are positive.
Introduce $\Phi_i^\prime =\tilde u_{ij}\Phi_j$
where ${\rm diag}\{\tilde\nu_1,\tilde\nu_2,\tilde\nu_3\}=\tilde u\,\tilde k\,\tilde u^\dagger$ 
and $\tilde k_{ij}$ is the matrix of coefficients of the  D-term 
(kinetic terms) in eq.(\ref{w0}). 
After rescaling $\Phi_i^\prime=\tilde \Phi_i/\sqrt{\tilde\nu_i}$ and with
$\tilde z_{kj}= (1/4)\,m\,\xi\,\tilde u_{j2}^* \,\tilde u_{k3}^*$, one  has
\medskip
\bea
\cL=\int d^4\theta 
\Big[
\tilde \Phi_1^{\dagger}\tilde \Phi_1
+
\tilde \Phi_2^{\dagger}\tilde\Phi_2
+
\tilde \Phi_3^{\dagger}\tilde \Phi_3
\Big]
+\Big\{
\int d^2\theta \Big[W\Big(\frac{\tilde u_{j1}^*\tilde\Phi_j}{\sqrt{\tilde\nu_j}}\Big)+
\frac{\tilde z_{kj}\,\tilde\Phi_j\tilde\Phi_k}{\sqrt{\tilde \nu_j\tilde \nu_k}} \Big]
+{\rm h.c.}
\Big\}
\eea

\medskip\noindent
The coefficients $\tilde u_{ij}$ and $\tilde z_{jk}$ are calculable 
and depend on  $\rho$, $\sigma$ and $\xi$.
 $\cL$  provides one possible
 UV theory\footnote{This is also a UV completion, since the UV theory is 
also renormalizable.} 
of the original Lagrangian eqs.(\ref{e0}), (\ref{e3}), 
is free of ghosts (under the above assumptions for $\rho,\sigma,\xi$),
 and recovers the initial effective Lagrangian. 

This method of identifying the UV theory can be 
extended to more complicated $K$ and $W$.
For this, we notice  that ghosts superfields emerge if
{\it kinetic mixing} of the superfields
or  bilinear derivative F-terms are present.
 In their absence,  even if higher derivative (interaction) terms 
exist, no ghost are generated, provided that 
the (scalar) fields present in the interactions are  not developing vev's.
If this is not true, 
the analysis  is complicated since interaction terms can contribute
to the two-point Green function and lead to ghost superfields.
In such case one could expand about the new ground state to identify 
their contributions to the  kinetic terms to see if any 
 ghosts superfields are generated.

\subsection{``Unfolding'' a higher-derivative ghost-free chiral theory}

\medskip
In this section we study  a different  theory with 
higher-derivatives that is  known  to be ghost-free,
since it does not introduce additional degrees of freedom 
(poles in the propagator)
 \cite{K1,K2,K3,K4}. Its Lagrangian is
\medskip
\begin{eqnarray}
 {\cal L}
&=& \int d^4 \theta \left\{  \Phi^{\dagger } \Phi  + 
\frac{\rho}{\Lambda^4}
D^{\alpha} \Phi D_{\alpha} \Phi  {\bar D}_{\dot \alpha} \Phi^{\dagger}
  {\bar D}^{\dot \alpha} \Phi^{\dagger}  \right\} 
\nonumber \\[4pt]
& =& \int d^4 \theta 
\,\Big\{  \Phi^{\dagger } \Phi +
\frac{\rho}{\Lambda^4}\,
\Big\vert\,\Phi D^2 \Phi - \frac{1}{2} D^2 \Phi^2\,\Big\vert^2
 \,\Big\}
+\cO(1/\Lambda^5) \ . \label{gf1}
\end{eqnarray}

\medskip\noindent
where the sign of $\rho$ is not yet fixed.
This effective action has  interactions of 
the form $|\partial \phi|^4$ and higher-order algebric
 terms for the chiral auxiliary field of the form $|F|^4$, 
but no dynamics for $F$ is generated.   
With the help of four chiral Lagrange (super)field multipliers $\Sigma_{1,2,3,4}$ one 
can rewrite  (\ref{gf1}) as
\medskip
 \begin{eqnarray}
 {\cal L}
&=& 
\int d^4 \theta \,\,
\Big\{  \Phi^{\dagger } \Phi  + \frac{\rho}{\Lambda^4}
 \Big\vert\, \Phi D^2 \Phi - \frac{1}{2} D^2 \Phi^2\,\Big\vert^2
\Big\} 
\nonumber \\[4pt]
& +& 
\Big\{
 \int d^2 \theta \,\, \big[ \Sigma_1 \big( m_1 \Sigma_2 - \epsilon
{\bar D}^2 \Phi^{\dagger }\big) + \Sigma_3 \big(m_2 \Sigma_4 -
(1/m_3)\,{\bar D}^2 \Phi^{\dagger 2 }\big) \big] + {\rm h.c.} 
\Big\}+\cO(1/\Lambda^5)\qquad
\eea

\medskip\noindent
Eliminating $\Sigma_{1,2,3,4}$ recovers the previous Lagrangian
while $m_{1,2,3}$ and $\epsilon$ 
are included for dimensional reasons. Further
\medskip
\bea
\cL\!\!& =&\!\! 
\!\int d^4 \theta\,\, \Big\{  \Phi^{\dagger } \Phi  + 4\, \epsilon\, (\Sigma_1 \Phi^{\dagger} + 
\Sigma_1^{\dagger} \Phi) +
\frac{\rho}{\Lambda^4}\,\,
\Big\vert 
\frac{m_1}{\epsilon} \,\Phi \,\Sigma_2^{\dagger} - \frac{ m_2\,m_3}{2}\, \Sigma_4^{\dagger} 
\,\Big\vert^2
\!\!\!+\! \frac{4}{m_3} (\Sigma_3 \Phi^{\dagger 2} + \Sigma_3^{\dagger} \Phi^2 ) \Big\} 
\nonumber\\[4pt]
&+&  \Big\{
 \int d^2 \theta \,\, \big(  m _1 \Sigma_1 \Sigma_2 +  m _2 \Sigma_3 \Sigma_4\big) 
+ {\rm h.c.} 
\Big\} +\cO(1/\Lambda^5).\qquad \label{gf2}
\end{eqnarray}

\medskip\noindent
In this ``unfolded'' formulation, the 
parameters $m_{1,2,3}$ and $\epsilon$ are 
not physical, but they become so in the UV theory (see later). 
This form of $\cL$ has no higher derivative terms  anymore.
Given the high dimension of the initial derivative operator, 
the terms in eq.(\ref{gf2}) that correspond to the initial operator,
although polynomial, have mass dimension larger than four.

Despite its appearance, the Lagrangian in eq.(\ref{gf2}) 
has no ghosts  since it is equivalent to the original (ghost-free) 
Lagrangian in eq.(\ref{gf1}). 
Indeed, for vanishing vev\footnote{If this is not the case, the analysis 
is more complicated.} $\langle \Phi \rangle = 0$, it is obvious  
$\Sigma_{2,3}$ have no dynamics and therefore only enforce  constraints on $\cL$. 
One of the constraints is that $\Sigma_1$ is a composite field, so
the apparent ``off-diagonal'' ghost-like kinetic term of  $\Sigma_1$ 
is actually a higher-order polynomial operator.

A UV theory of eq.(\ref{gf2}) is found similar to the previous
examples by adding the missing kinetic terms 
\medskip
\begin{eqnarray}
 \delta {\cal L}_{\rm kin.} 
&= & \int d^4 \theta \ (  \Sigma_1^{\dagger } \Sigma_1
+  \Sigma_2^{\dagger } \Sigma_2 + \Sigma_3^{\dagger } \Sigma_3 ) \ , 
\nonumber \\[6pt]
{\cal L}_{\rm UV} & =&  \ {\cal L}
 + \delta {\cal L}_{\rm kin.}  \ . 
 \label{gf3}
\end{eqnarray} 

\medskip\noindent
The resulting UV Lagrangian is manifestly ghost-free for 
$\epsilon < \frac{1}{4}$ and
$\rho > 0$, which is the appropriate sign coming from general considerations 
\cite{nima}. 
Notice that in the UV Lagrangian of eq.(\ref{gf3}) all parameters are physical, 
so the theory has now several mass scales. By integrating out the  massive
 fields $\Sigma_i$ one  finds the  corresponding effective action.

One can ask if this UV theory also generates additional effective operators 
beyond the original one in (\ref{gf1}) of dimensions lower or equal to its dimension.
The field equations  of the massive fields determine their solution:
\medskip
\begin{eqnarray}
\Sigma_1 &=& \frac{4 \epsilon}{m_1^2} \Big (1 + \frac{\Box}{m_1^2}\Big) 
\Box \Phi 
- \frac{\rho}{4 \epsilon\,\Lambda^4} {\bar D}^2 (\Phi^{\dagger} D^{\alpha}
 \Phi D_{\alpha} \Phi) + \cdots\ , 
\nonumber \\[7pt]
\Sigma_2 &=& \frac{\epsilon}{m_1} \Big(1 + \frac{\Box}{m_1^2}\Big)
 {\bar D}^2 \Phi^{\dagger} + \cdots \ ,
\nonumber \\[7pt]
\Sigma_3 &=& \frac{2 \rho\,m_3}{\Lambda^4} (\partial \Phi)^2 + \cdots \ , 
\nonumber \\[7pt]
\Sigma_4 &=&  \frac{1}{m_2 m_3} {\bar D}^2 \Phi^{\dagger 2} +
\frac{\rho\,m_3}{2\, m_2 \,\,\Lambda^4} {\bar D}^2 (\partial \Phi^{\dagger})^2 + \cdots \ , 
\end{eqnarray}

\medskip\noindent
where the dots  stand for terms that contribute to operators 
of dimension higher than eight.
Substituting this solution back in $\cL_{\rm UV}$ of eq.(\ref{gf3}) one
 finds a low energy,  effective  Lagrangian below
\medskip
\begin{equation}
 {\cal L}_{\rm eff} = \int d^4 \theta \left\{  \Phi^{\dagger } \Phi  + 
\frac{16 \epsilon^2}{m_1^2} \Phi^{\dagger} \Box \Phi + \frac{16 \epsilon^2}{m_1^4} \Box \Phi^{\dagger} \Box \Phi + \frac{\rho}{\Lambda^4}
D^{\alpha} \Phi D_{\alpha} \Phi  {\bar D}_{\dot \alpha} \Phi^{\dagger}  {\bar D}^{\dot \alpha} \Phi^{\dagger} + \cdots \right\} \ . 
\end{equation} 

\medskip\noindent
We thus recovered our initial operator, the last term above, suppressed by $\Lambda$.
The other two operators are suppressed by another mass scale $\frac{m_1}{\epsilon}$.
One can in principle arrange that  $\Lambda \ll m_1/\epsilon$ by a suitable choice for 
(dimensionless) $\epsilon$. In this case our  operator of interest in eq.(\ref{gf1}) is the 
leading one generated  at low energy.  

As seen above, consistency of the UV theory we found demands that $\rho=1$.

\section{Effective operators for a linear multiplet}
\label{linear}

Another important supersymmetric multiplet,  notably in string models, 
is a real linear multiplet $L$ \cite{linear}, containing a real scalar field 
$c$, a fermion $\psi$ and an antisymmetric tensor $b_{\mu\nu}$. 
The multiplet satisfies the constraints
\bea
D^2 L = {\oD}^2 L = 0 \ ,   \label{l1}
\eea
with a solution
\bea
L = D^{\alpha} \cZ_{\alpha} + {\overline D}_{\dot \alpha} {\overline \cZ}^{\dot \alpha}
\ ,   \label{l2}
\eea

\medskip\noindent
where $\cZ_{\alpha}$ is a chiral spinor superfield. 
The superspace
expansion in the rigid case is
\medskip
\bea
L = c + \theta \psi + {\bar \theta} {\bar \psi} 
-
 \theta \sigma^\mu {\bar \theta}
\epsilon_{\mu\nu \rho\sigma} \partial^\nu b^{\rho\sigma} 
+ 
\frac{i}{2} {\theta}^2 {\bar \theta} {\bar \sigma}^\mu \partial_\mu
{\psi} 
-
 \frac{i}{2} {\bar \theta}^2 \theta \sigma^\mu \partial_\mu {\bar \psi}
+\frac{1}{4} \theta^2 {\bar \theta}^2 \Box c \ . \label{l3}
\eea

\medskip\noindent
The massless action for a linear multiplet has the gauge symmetry
\begin{equation}
\cZ_{\alpha} \to \cZ_{\alpha} - i W_{\alpha} \ , 
\end{equation}
if $W_{\alpha}$ is a gauge superfield strength satisfying  $D^{\alpha } W_{\alpha} = 
{\overline D}_{\dot \alpha } {\overline W}^{\dot \alpha}$. 
The free action of a massive linear multiplet is \cite{linear-multiplet-sign}
\bea
{\cal L} = \int d^4 \theta \ \Big( - \frac{1}{2} L^2\Big) 
+ \Big\{ \int d^2 \theta \ \frac{-\Lambda^2}{2} \cZ^{\alpha}
\cZ_{\alpha} + {\rm h.c.} \Big\}  \ . 
\eea

\subsection{``Unfolding'' effective operators in the linear multiplet sector}

In applications one can encounter an effective operator of the type shown below
acting in the linear multiplet sector 
\medskip
\bea
{\cal L}_0 = \int d^4 \theta\,\, \Big[\, - \frac{1}{2} L^2 + \frac{\rho}{\Lambda^2}
 L \Box L  \, \Big]  +\cO(1/\Lambda^3)
\label{l5}
\eea

\medskip\noindent
where $\rho=\pm 1$ was introduced to allow either sign for the last term.
 $\cL_0$  can  be written as
\medskip
\bea
{\cal L}_0 = \int d^4 \theta \,\,
\Big[ - \frac{1}{2} L^2 +
 \frac{\rho}{8 \,\Lambda^2} \,
L\, D^{\alpha} {\bar D}^2 D_{\alpha} \,L  
\Big]+\cO(1/\Lambda^3) \ , \label{l6} 
\eea

\medskip\noindent
by using the identity 
$\oD_{\dot\alpha} D^2 \oD^{\dot\alpha}=
D^{\alpha} {\overline D}^2 D_{\alpha} = \frac{1}{2} \{ D^2, {\bar D}^2 \} + 8 \ \Box $. 
The above effective operator can be ``unfolded'' by using our
experience so far\footnote{
According to our method the constraint 
$ L' = D^{\alpha} \cZ'_{\alpha} + {\oD}_{\dot \alpha} {{\overline\cZ'}}^{\dot \alpha} $
should be imposed via a Lagrange multiplier superfield. We 
impose this constraint 
directly when using the eq of motion
of $\cZ^\prime$  instead of (constrained) $L^\prime$.}.
The result can be guessed directly, by starting with
\medskip
\bea
{\cal L} = \int d^4 \theta \Big[ - \frac{1}{2} L^2 + a\, L L' \Big] +
\Big\{ \int d^2 \theta \ \frac{-\Lambda^2}{2} \cZ'^{\alpha} \cZ'_{\alpha} + {\rm h.c.} \Big\}
 \ , \label{l7}   
\eea 

\medskip\noindent
with a constraint $ L' = D^{\alpha} \cZ'_{\alpha} + {\oD}_{\dot \alpha} 
{{\overline\cZ'}}^{\dot \alpha} $; here
$L'$ is a massive linear multiplet.
$a$ is a real dimensionless numerical coefficient to be identified shortly.
The above Lagrangian has one standard linear multiplet ($L$) and
an additional,  ghost linear multiplet 
since the determinant of the kinetic terms matrix is negative $\propto -a^2<0$.
One can eliminate
 the massive linear multiplet $L'$ via its equation of motion that is
actually obtained from that for the unconstrained (independent) field  $\cZ^\prime$
\medskip
\bea
\cZ'_{\alpha} = \frac{a}{4 \Lambda^2} {\oD}^2 D_{\alpha} L \qquad  \Ra\qquad
L' =   \frac{a}{4 \Lambda^2}\,\, \big( D^{\alpha}{\oD}^2 D_{\alpha} + {\oD}_{\dot \alpha}{D}^2 
{\oD}^{\dot \alpha} \big) L \ . \label{l8}
\eea

\medskip\noindent
The effective Lagrangian after eliminating $L^\prime$  becomes
\medskip
\bea
\cL & =&
  \int d^4 \theta \Big[
 - \frac{1}{2}\, L^2 + \frac{2 \, a^2}{\Lambda^2}\, L \,\Box\, L   \Big] \ , \label{l9} 
\eea

\medskip\noindent
This is identical to eq.(\ref{l5}) provided that  
$a^2=\rho/2$ which has a solution for $\rho=+1$ only.
With this value, $\cL$ of eq.(\ref{l7}) is an ``unfolded'' version of initial
 $\cL_0$ of eq.(\ref{l5}) since it has no higher dimensional operators, 
but it contains an additional ghost-like superfield.

\subsection{Ghost-free UV theory for the linear multiplet case}

A natural UV theory of the effective Lagrangian above is to add to 
eq.(\ref{l7}) a kinetic  term for the massive linear multiplet $L'$ 
\medskip
\begin{equation}
{\cal L} = \int d^4 \theta \left[ - \frac{1}{2} L^2 - \frac{1}{2} L'^2 + a L L' \right] +
\left\{ \int d^2 \theta \ \frac{-\Lambda^2}{2} \cZ'^{\alpha} \cZ'_{\alpha} + {\rm h.c.} \right\}
 \ , \label{l10}   
\end{equation}

\medskip\noindent
$\cL$ is ghost-free (if $a^2<1$) and  recovers the Lagrangian ${\cal L}_0$  plus higher 
derivative operators.

Other UV theories are possible. 
Consider for example the coupling of the massless linear multiplet
$L$ to a massive vector multiplet $V$
\medskip
\bea
{\cal L}_1 = \int d^4 \theta\,\, \Big(
\, - \frac{1}{2} L^2 - M^\prime L V + \frac{M^{2}}{2} V^2\,\, \Big)
+ \Big\{
\int d^2 \theta \ \frac{1}{4} W^{\alpha} W_{\alpha} + {\rm h.c.} 
\Big\} \ . \label{l11}
\eea

\medskip\noindent
The field equation of the massive vector multiplet leads to
\medskip
\bea
V = \frac{M^\prime}{M^{2} + (1/4) \, D^{\alpha} {\bar D}^2  D_{\alpha}} L =
\frac{M^\prime}{M^{2} + 2 \Box } L  \ , \label{l12}
\eea

\medskip\noindent
which, when inserted back into eq.(\ref{l11}), leads to
\medskip
\bea
{\cal L}_{1,{\rm eff}} = \int d^4 \theta 
\,\,
\Big[\, - \frac{1}{2} L^2 - \frac{M^{\prime 2}}{2}
 L  \frac{1}{M^2 +  2 \Box } L \, \Big]  
\label{l13}
\eea

\medskip\noindent
This Lagrangian
is equivalent to the original effective Lagrangian  ${\cal L}_{0}$ of eq.(\ref{l5})
after an expansion  in $\Box/M^\prime$ and 
 a wave function renormalization of $L\ra L/(1+M^{'\,2}/M^2)^{1/2}$ and with the identification
 $\Lambda\equiv M (1+M^2/M^{' 2})^{1/2}$. 
Like  in the previous example,   $\rho=+1$.

\section{Effective operators in the gauge sector}
\label{s4}

\subsection{``Unfolding'' the effective operators in the gauge sector }

The above analysis can be extended  to cases when such
effective operators  are present in the gauge sector. 
Without restriction to generality, we consider an Abelian case with\footnote{
We denote the gauge field strength by $W^\alpha$, not to be confused with the
superpotential $W$.}
\medskip
\bea\label{Fbox}
\cL=\int d^4\theta \, K(\Phi^i, e^V, \Phi^\dagger_j, ....)
 +\Big\{\int d^2\theta\,\Big[\,\frac{1}{4} W^\alpha W_\alpha
 -  \frac{\rho}{\Lambda^2}\,\, W^\alpha\Box W_\alpha\Big] + {\rm h.c.}\Big\}
+\cO(1/\Lambda^3)
\eea

\medskip
\noindent
where $\rho=\pm 1$. 
With $W_\alpha =-(1/4)\,\overline D^2 D_\alpha V$ then
\medskip
\bea
 \delta \cL 
= - \frac{\rho}{\Lambda^2} \int d^2\theta     \,\, W^\alpha\Box W_\alpha
 = - 
\frac{\rho}{4 \Lambda^2} \int d^4\theta\,\, W^\alpha D^2 \,W_\alpha 
=
\frac{-\rho}{2 \Lambda^2}\int d^4\theta\,(D^\alpha W_\alpha)^2\quad
\eea

\medskip\noindent
where we used that $D^2\epsilon^{\gamma\alpha}=-2 D^\gamma D^\alpha$. Then
\medskip
\bea
\cL\!=\!\!
\int\!\! d^4\theta \Big[
K(\Phi^i, e^V\!, \Phi^\dagger_j)
\!-\!
\frac{\rho}{2\Lambda^2} \big[\,(D^\alpha W_\alpha)^2
\!+ \!
(\overline D_{\dot \alpha} \overline W^{\dot \alpha})^2\big]\,
\Big]
\!\! +\!
\Big\{\!\!\int\! d^2\theta
\frac{1}{4}\, W^\alpha W_\alpha\! +\! {\rm h.c.}\Big\}
\!\!+\!\!\cO\Big[\frac{1}{\Lambda^3}\Big]
\eea

\medskip\noindent
As before, we introduce an auxiliary (real) superfield $V^\prime$ which  enables
 us to remove the higher dimensional (derivative) operator
 via a constraint
\medskip
\bea
\label{co}
D^\alpha W_\alpha=m^2\,V^\prime
\eea
Here  $m$ is a small arbitrary
 scale of the theory which is set to 0 at the end of the 
calculation. We implement
the constraint using a Lagrangian multiplier which is  a real superfield 
$\Sigma$,  as shown below:  
\medskip
\bea
\cL &=&
\int  d^4\theta \Big[
K(\Phi^i, e^V, \Phi^\dagger_j)\,
- \frac{\rho\, m^4}{\Lambda^2} \,V^{\prime 2}
+2\,
\Sigma \,(D^\alpha W_\alpha - m^2\, V^\prime) \Big]
\nonumber\\
& + &
\Big\{\int 
 d^2\theta\,\,\frac{1}{4}\, W^\alpha W_\alpha + {\rm h.c.}\Big\}  +\cO(1/\Lambda^3)
\label{hh}
\eea

\medskip\noindent
Using the Lagrangian in eq.(\ref{hh}), 
the eq of motion for $\Sigma$ reproduces the constraint eq.(\ref{co}).
$V^\prime$
can be eliminated (integrated out exactly) since its equations of motion are algebric,
so $\cL$ becomes:
\medskip
\bea
\cL&=&
\int d^4\theta \,\Big[
K(\Phi^i, e^V, \Phi^\dagger_j)
+\rho\,
\Lambda^2\,\Sigma^2
\Big]
\nonumber\\[4pt]
&+&
\Big\{\int d^2\theta\,
\,\,\Big[\,\,\frac{1}{4}\,
 W^\alpha W_\alpha 
-
\,W^\alpha(\Sigma) \,W_\alpha
\Big]
+{\rm  h.c.}\Big\}
+\cO(1/\Lambda^3)
\label{rr}\qquad
\eea
where
$ W^\alpha(\Sigma)=-\frac{1}{4} \overline D^2 D^\alpha \Sigma$.
We obtained a second-order theory with renormalizable interactions (by power counting)
with the original vector superfield $V$ and an additional massive one $\Sigma$.
The new field $\Sigma$ is a ghost (vector) superfield since
the determinant of the  kinetic terms is negative.
The gauge kinetic mixing can be diagonalised by a rotation 
and an appropriate rescaling. 
Finally, one can also integrate out $\Sigma$
using the above $\cL$ to obtain an effective operator of
second order, as done for the matter sector, eqs.(\ref{eer}), (\ref{eer2}).
The result depends on the structure assumed for $K$ (for example one can consider the simplest
case $\Phi^\dagger e^V\Phi$, etc).

\subsection{A ghost-free  UV completion in the gauge sector}
\label{s5}

Can we can find in this case a simple, ghost-free (renormalizable) 
UV theory  of $\cL$ in eq.(\ref{Fbox}), (\ref{rr})?
 From the  matter sector we 
know that simply adding a positive   kinetic term, in this case
 for $\Sigma$ in eq.(\ref{rr}), is the way to proceed. 
We thus add
\medskip
\bea
\delta \cL=\int d^2\theta\, \,\delta\,\, W^\alpha(\Sigma)\,W_\alpha(\Sigma)+{\rm h.c.}
\eea

\medskip\noindent
$\delta\cL$ together with $\cL$ of eq.(\ref{rr}) can be brought to a diagonal
basis after a suitable rotation applied to $V, \Sigma$. Then
one chooses values for $\delta$  so that there are no ghost vector 
superfields in $\cL+\delta\cL$.
This new Lagrangian gives a UV theory that is also renormalizable
(UV completion).

\subsection{Application: an example  generating  Dirac gaugino masses}
\label{dirac}

As discussed, not all theories with $\cL$ 
having powers of superderivatives generate ghosts, 
if these are present in interactions. We  construct another
example in the following. Start with a model
\medskip
\bea\label{kkj}
\cL\!=\!\!
\int\! d^4\theta\,\Big[\,
\frac{\Lambda^{\prime 2}}{2}\,V^{\prime 2}
+
X^\dagger e^{V^\prime}X
+
\Phi^\dagger\Phi
\Big]
\!+\!
\Big\{\!\!
\int d^2\theta\,\Big[\,\,\frac14\, W^\alpha\,W_\alpha
\!+\!\frac{1}{\Lambda}\,W^{\prime \alpha}\,W_\alpha\Phi
\!+\!
W(\Phi)\Big]
\!\!+\!\!{\rm h.c.}\!\!\Big\}
\eea

\medskip\noindent
where $\Lambda^\prime$ is large, comparable to $\Lambda$.
We have a massive gauge field $V^\prime$ of field strength
$W^\prime_\alpha$, a field $X$ charged under it, a gauge kinetic term 
 for $V$ and an interaction term with $\Phi$ neutral under $V, V^\prime$.
If $\Phi$ has a scalar component with non-zero vev, then we would 
have kinetic mixing $W\,W^\prime$
and this would induce the presence of a ghost (for $W^\prime$). However, 
assume  that the scalar component of $\Phi$ has a  vanishing vev, 
ensured by a suitable choice of the superpotential  $W(\Phi)$.
Then the field equation for $V^\prime$ gives
$V^\prime(\Lambda^{\prime 2}
+X^\dagger X)=1/\Lambda\,\big[D^\alpha \big(W_\alpha\Phi\big)
+{\rm h.c.}
\big] -X^\dagger X$.
Using this in $\cL$, we obtain an effective Lagrangian 
\medskip
\bea
\cL_{\rm eff}&\supset &
\int d^4\theta \Big[
\Phi^\dagger\Phi+X^\dagger X
-\frac{1}{2\,\Lambda^{\prime 2}}
\Big( 1/\Lambda\,\,\big[
D^\alpha \big(W_\alpha\Phi\big)
+{\rm h.c.}
\big]
\!-X^\dagger X\Big)^2 \Big]
\nonumber\\
&+&
\int d^2\theta\, \Big[\,\,\frac14\,W^\alpha W_\alpha+W(\Phi)\Big]+{\rm h.c.}
\eea

\medskip\noindent
up to additional terms (not shown) suppressed by extra powers $\Lambda^\prime$.
If we identify  the chiral superfield $X$ with the spurion of supersymmetry breaking,
then $\cL_{\rm eff}$ contains a Dirac mass term \cite{nelson1,nelson2}
\bea
\cL_{\rm eff}\supset \frac{1}{\Lambda^{\prime 2}\,\Lambda}\int d^4\theta
\Big[ X^\dagger X\, D^\alpha\big(W_\alpha\,\Phi\big) +{\rm h.c.}\Big]
\eea
This is seen  by considering the fermionic component of $\Phi$ and the gaugino $\lambda$
in $W_\alpha$ giving  a mass term for $\lambda\,\psi$ 
where $\psi$ is the Weyl fermion of $\Phi$.
Finally, a  UV theory 
 of Lagrangian (\ref{kkj})  is obtained by adding there an F-term 
\medskip
\bea
\delta\cL=\frac14\int d^2\theta \,W^{\prime\alpha}\,W_{\alpha}^\prime +{\rm h.c.}
\eea

\medskip\noindent
We obtain in this way a two-derivative, ghost free 
Lagrangian that generates in the low energy the 
Dirac gaugino mass term.

\subsection{Supersymmetric Euler-Heisenberg Lagrangian}

Another interesting case is that of a supersymmetric generalisation of the
Euler-Heisenberg Lagrangian  which is a  higher-derivative gauge theory
that can be  ghost-free\footnote{
Here we ``unfold'' this operator  at the classical level.
However this operator 
 can also be generated by loop corrections in a standard gauge theory.
For studies of this operator in supergravity, see e.g. \cite{sergio}.}.
It is given by
\medskip
\begin{equation}
{\cal L} \!= \!
\int d^4 \theta \ \frac{\rho}{\Lambda^4}
\ W^{\alpha} W_{\alpha} {\overline W}_{\dot \alpha} {\overline W}^{\dot \alpha}
+
\Big\{ \int d^2 \theta \,\,\frac{1}{4} W^{\alpha} W_{\alpha} + {\rm h.c.} \Big\}
\label{eh1}
\end{equation}

\medskip\noindent
where $\rho=\pm1$ accounts for possible signs of this operator.
$\cL$ contains the gauge field term
\medskip
\begin{equation}
{\cal L} = - \frac{1}{4} F_{\mu \nu} F^{\mu \nu}+ \frac{\rho}{4 \,\Lambda^4} 
\left[ (F_{\mu \nu} F^{\mu \nu})^2 + 
(F_{\mu \nu} {\tilde F}^{\mu \nu})^2 \right] + \cdots \ , \label{eh2}
\end{equation}

\medskip\noindent
One must have $\rho > 0$  according to the constraints discussed in \cite{nima}.
Similar to the chiral superfields example discussed earlier, 
 $\cL$ can be re-formulated  as (``unfolded'' into) a second-order theory. 
The idea is to introduce a constraint that replaces
 a pair $W^\alpha W_\alpha$ of the D-term by a chiral 
superfield so this term  becomes  quadratic (in $S$) 
 and thus ``removes'' the extra derivatives.
We thus introduce two additional chiral superfields $\Phi$ and $S$, 
according to
\medskip
\begin{equation}
{\cal L} = \int d^4 \theta \ S^{\dagger } S +  
\left\{
 \int d^2 \theta \left[ \frac{1}{4} W^{\alpha} W_{\alpha}
  + M \,\Phi (S - \epsilon \, W^{\alpha} W_{\alpha})  \right] +  {\rm h.c.} \right\},
 \label{eh3}
\end{equation}

\medskip\noindent
where $\Phi$ is the Lagrange multiplier (chiral superfield) 
that implements the constraint and $M$ is an arbitrary mass scale. 
Notice that,
up to total derivatives, there is a continuous shift symmetry 
$\Phi \to \Phi + i \alpha$, where $\alpha$ is a real parameter.  
After integrating out $S$ and $\Phi$ one recovers
$\cL$ of (\ref{eh1}) provided that $\rho/\Lambda^4 = \epsilon^2$. 
It is interesting that this second-order formulation
 of the  starting $\cL$   is actually consistent with 
the  condition $\rho>0$ \cite{nima}. 

The situation is however more subtle because the
 chiral operator  $W^{\alpha} W_{\alpha}$ is constrained by 
$ W^{\alpha} W_{\alpha} = \frac{1}{2} {\oD}^2 \Omega$, where the real 
superfield $\Omega$ is the Chern-Simons  superfield  \footnote{
In the Abelian case, it is given explicitly by 
$\Omega = - \frac{1}{2} \left( D^{\alpha} W_{\alpha} + 
{\bar D}_{\dot\alpha} {\bar W}^{\dot \alpha} + V D^{\alpha} W_{\alpha} \right). $} \cite{g2,g3}. 
Accordingly, the field $S$  in eq.(\ref{eh3}) is not an independent
chiral superfield, but the field strength of a three-form superfield 
$S = - \frac{1}{4} {\oD}^2 U$,
where $U$ is a dimensionless superfield\footnote{This is similar to 
the dilaton case, for example.}.

Let us briefly review some details regarding the 
 real three-form multiplet $U$. This is defined in superspace by 
\medskip
\begin{eqnarray}
&& U  = {\overline U} = B + i (\theta \chi - {\bar \theta} {\bar \chi}) + \theta^2 {\bar s} +
 {\bar \theta}^2 { s} + \frac{1}{3} \theta \sigma^m {\bar \theta} \epsilon_{mnpq} C^{npq}
+ \nonumber \\ 
&& \theta^2 {\bar \theta}  (\sqrt{2} {\bar \lambda} + \frac{1}{2} {\bar \sigma}^m \partial_m \chi)
+ {\bar \theta}^2 \theta  (\sqrt{2} {\lambda} - \frac{1}{2} {\sigma}^m \partial_m {\bar \chi})
+ \theta^2 {\bar \theta}^2 (D+\frac{1}{4} \Box B) \ .
\end{eqnarray}

\medskip\noindent
The difference between a three-form multiplet $U$ and a regular vector 
superfield $V$ is the replacement of the vector potential $V^m$ by the three-form $C^{npq}$. 
To find the appropriate kinetic terms, the analog of the chiral field strength superfield 
$W_{\alpha}$ for a vector multiplet is the chiral superfield \cite{g1,g2}
\medskip
\begin{equation}
S = - \frac{1}{4} {\oD}^2 U \quad , \quad S (y,\theta) = s + \sqrt{2} \theta \lambda +
\theta^2 (D + i F) \ , 
\end{equation}

\medskip\noindent
with $F$ defined by  $F = \frac{1}{4 !} \epsilon_{mnpq} F^{mnpq}$, where $F^{mnpq}$ is the
 four-form field strength. While
 the massless three-form multiplet $U$ has only two 
propagating bosonic ($s$) and two fermionic ($\lambda$) degrees of freedom, a massive
 three-form multiplet has four bosonic and fermionic degrees of freedom. Notice also
 that a massless three-form multiplet has the gauge invariance $U \to U - L$, where 
$L$ is a linear multiplet. This symmetry is broken by a mass term, similarly 
to the case of a standard vector multiplet $V$.   References \cite{g1,g2} contain
detailed explanations about the three-form multiplet.

Returning to eq.(\ref{eh3}), there 
is a dual formulation of this equation, 
which can be found by starting from the master Lagrangian
\medskip
\begin{equation}
{\cal L} = \int d^4 \theta \ \left[ S^{\dagger } S  + M^2 Q (U-L + 2\,
 \,\epsilon \,\,\Omega) \right]
+  \left\{ \int d^2 \theta \frac{1}{4} W^{\alpha} W_{\alpha} + {\rm h.c.} \right\}
\ , \label{eh5}
\end{equation}

\medskip\noindent
where $Q$ is a real superfield and $L$ a (dimensionless) linear multiplet. 
The equation of motion for  $L$ gives $M Q = \Phi + \Phi^{\dagger}$, which when used 
into eq.(\ref{eh5}) recovers the action   of eq.(\ref{eh3}). 
Alternatively,  eliminating $Q$ one obtains
\medskip
\begin{equation}
U = L - 2\, \epsilon\, \Omega \quad , \quad 
S = \epsilon \, W^{\alpha} W_{\alpha} \ , \label{eh6}
\end{equation}

\medskip\noindent
which inserted into (\ref{eh5}) does  recover
 the starting Lagrangian of  eq.(\ref{eh1}) (with $\epsilon^2=\rho/\Lambda^4>0$).

Finding the UV theory of  the Euler-Heisenberg Lagrangian is now done
by adding the missing kinetic term in the chiral formulation 
\medskip
\begin{equation}
{\cal L}_{\rm UV} = \int d^4 \theta 
\Big[
S^{\dagger } S +   \frac{1}{2}(\Phi +\Phi^{\dagger})^2 \Big] 
+  
\Big\{
 \int d^2 \theta \Big[\, \frac{1}{4}\, W^{\alpha} W_{\alpha}
  + M \Phi (S - \epsilon  W^{\alpha} W_{\alpha})  \Big]
 +  {\rm h.c.} \Big\} \ . 
\label{eh7}
\end{equation}

\medskip\noindent
The shift symmetric kinetic term for $\Phi$ is 
equivalent to a standard canonical one in rigid SUSY, 
but is different in the supergravity version.
The dual UV Lagrangian is  found
 by introducing a vector multiplet as above and 
eliminating it out via its equation of motion. The result is
\medskip
 \begin{equation}
{\cal L} = \int d^4 \theta \ \Big[
 S^{\dagger } S  - \frac{M^2}{2}\, (U-L + 2 \epsilon\, \Omega)^2 \Big]
+  \left\{
 \int d^2 \theta \, \frac{1}{4} \, W^{\alpha} W_{\alpha} + {\rm h.c.} \right\}
\ , \label{eh8}
\end{equation}

\medskip\noindent
and contains a massive three-form multiplet. 
The action in eq.(\ref{eh8}) is fully gauge invariant.
In particular, under $U(1)$ gauge transformations and in 
the unitary gauge $L=0$, the relevant transformations are
\begin{equation}
\delta V = \Theta + {\bar \Theta} 
\quad , \quad \delta U = 
\epsilon \left[ D^{\alpha} (\Theta W_{\alpha}) + {\bar D}_{\dot \alpha} 
(\Theta^{\dagger } \overline{ W^{\dot \alpha}})  \right]
\ . \label{eh9}  
\end{equation}  
This discussion 
neglected subtleties related to the existence of boundary terms in the action. 
While  they are important to obtain a fully consistent Lagrangian
 (see e.g. \cite{g4}), they are not relevant for the above discussion. 

It is interesting to note that the Lagrangian in eq.(\ref{eh8})
(with the gauge field set to zero) is that of the chaotic inflationary
 model described in the last reference in \cite{g4}.

\section{More general cases}
\label{s7}

The method of  superfield constraints
that we introduced can be generalised  to more arbitrary $K$, $W$
which have as arguments {\it chiral} functions
\medskip
\begin{equation}
\cL=\!
\int\! d^4\theta\,\, K(\Phi_j,\Phi_j^\dagger, \overline D^2\Phi_j^\dagger,  D^2\Phi_j, 
D^2\overline D^2 \Phi_k^\dagger\cdots)
+\Big\{\!
\int\! d^2\theta \,\, W(\Phi_j,\overline D^2\Phi_j^\dagger,\overline D^2 D^2\Phi_k\cdots)
+{\rm h.c.}\Big\}\\
\end{equation}

\medskip\noindent
Note the dependence on  the superderivatives of many fields $\Phi_j$. 
The dots stand for powers of such  superderivatives that may also be present. 
All higher dimensional terms 
are suppressed by appropriate powers of a high scale 
(the Kahler curvature tensor). This  action is difficult to compute and
investigate in component fields.
However, one can introduce constraints
$\overline D^2\Phi_i^\dagger=m\,\tilde\Phi_{i}$,  $i=1,2....$, and similar for 
the other, higher order superderivatives that may be present, by
using an iterative procedure (as done in the Appendix).
These constraints are then added to the original $\cL$ as F-terms of type
\bea
\frac{m}{\Lambda}
\int d^2\theta\, \sum_{j\geq 1}\,\,(\overline D^2\Phi_j^\dagger-
m\,\tilde\Phi_j)\, \Sigma_j+{\rm h.c.}
\eea
 with $\Sigma_j$ the Lagrange multipliers chiral superfields
whose eqs of motion recover the constraints.  As in previous cases,
the coefficient in front of the integral
is {useful} to ensure the constraint is vanishing in the limit
$\Lambda\ra\infty$ and that the kinetic mixing is under control.
Similar considerations apply for 
the vector (gauge) or linear multiplet  sectors.
In this way one obtains  a $K$ and $W$ that only depend on 
the superfields $\chi_k=\{\Phi_i,\tilde\Phi_j,\Sigma_l\}$ but not on the superderivatives. 
With $K=K(\chi_k,\chi_k^\dagger)$ and $W(\chi_k)$,
 the Grassmann integrals in $\cL$ can then  be performed as  
for the non-linear sigma-model by using a (Grassmann space) Taylor  expansion, to 
find the component fields action \cite{WB}. This will be
a second order theory,  but with additional fields (that can then 
be eliminated by the field equations, if massive enough).

The ``unfolding'' method of superfield constraints 
applied to higher order terms in momentum expansion of the Lagrangian
can also be applied to other general cases. One may be  
interested in cases when the momentum is closer to the effective cutoff,
when even higher order terms (beyond those considered here) are relevant.
In the Appendix we present  the extreme case of including  {\it all orders} 
in such an expansion and apply our  method.
To illustrate it, we simplify the analysis and  use instead 
a {\it known UV-complete} (renormalizable) theory, eq.(\ref{hhh}) 
 and integrate a massive state of mass $\Lambda$ to all orders 
in $\partial/\Lambda$, to generate such a momentum expansion.
The new theory, eq.(\ref{tw})  is non-local \cite{georgi}
 equivalent to the initial one, eq.(\ref{hhh}).
We  then use an infinite set of superfield constraints, eqs.(\ref{qq}) to
eliminate all powers of the superderivatives.
This shows  that even in this case  there still exists an ``unfolding'' 
formulation of the higher order theory,  in which the infinite series (in 
powers of $\partial/\Lambda$) of effective  operators
is replaced by  polynomial (quadratic) terms in superfields, plus
 extra superfields that are massive (mass of order $\Lambda$), 
see eqs.(\ref{fff}), (\ref{f44}). 
Half of these massive superfields are ghost-like and half
 of them are particle-like. 
 Truncating this theory to a given number of superfields is
equivalent to truncating the initial Lagrangian to a given power of $\partial/\Lambda$.
In this ``truncated'' theory,  one can integrate out these massive 
fields to obtain a Lagrangian with new effective operators  polynomial in superfields, as in eq.(\ref{eer}), (\ref{eer2}). 
While this example  is very simple, it shows that our method can be
extended to higher or all orders in momentum expansion
by using an appropriate set of superfield constraints and their iteration.

\section{Conclusions}
\label{c}

Effective field theories contain a series in momentum expansion
that includes a special class of higher dimensional operators. 
These are generated by  classical integration of massive states
even in renormalizable, UV complete
 theories (and also by quantum  corrections of compactification).
The result of such integration is in general truncated to 
a given order in momentum expansion $\partial/\Lambda$
($\Lambda$ is the effective cutoff). 
This leads to an effective theory with  higher  derivative operators 
that can induce the presence of (super)fields of negative 
kinetic terms (ghosts). 
Contrary to a common perception, 
there is nothing ``pathological'' about their presence here. 
They are just artefacts of the expansion in $\partial/\Lambda$
obtained after integrating out massive states (of mass $\sim\Lambda$)
 of the UV theory.
Their presence and that of the corresponding (derivative) operators
is not problematic and we showed how to treat them. 
Finally, keeping all orders in the 
momentum expansion gives a non-local theory equivalent (classically)
to the initial, fundamental theory that generated the effective
operators in the low energy.

In a manifestly supersymmetric approach and using superfield constraints,
we showed how such effective operators acting on chiral,  vector and  linear superfields
can be ``unfolded'' into ``standard'' operators, polynomial in superfields, 
in the order of truncation in $\partial/\Lambda$ considered.
To see this, consider the case of chiral superfields. In such case
one replaced the superderivatives $\overline D^2\Phi_0^\dagger$ by
new chiral superfields $m\chi=\overline D^2\Phi_0^\dagger$ where $m$ is a small,
arbitrary mass introduced for dimensional reasons. This chiral
superfield constraint was enforced with a Lagrange multiplier
chiral superfield and in this way all superderivatives are eliminated.
The method can be applied and iterated to higher orders in $D, \oD$ 
(and also to non-derivative  operators).
After this  ``unfolding'' there are no superderivatives left
but only terms polynomial in superfields and 
additional  superfields (some of which are ghost-like) of
a mass of order $\Lambda$. 
These can be integrated out to obtain a ghost-free low energy effective
action that is  polynomial in superfields and corresponding to   the 
order in  $\partial/\Lambda$ considered.
The action so obtained can then be used for phenomenology.
This procedure can be repeated to higher orders in $\partial/\Lambda$ for
improved accuracy.  The method was then  applied to cases when 
superderivative operators  act in  the gauge and linear multiplets sectors, with 
similar results.

The ``unfolding'' method of  superfields constraints  helps one  identify 
(two-derivative and ghost-free) UV theories that generate at low-energy the 
effective operators considered.
We applied this method to the case when $\Box$-operators acted on
 chiral, vector or linear superfields, and using their ``unfolded'' formulation
we identified the associated UV theory (not necessarily unique). 
In these examples the initial effective operators were 
{\it quadratic} in fields or gauge fields strengths.

Further examples  were provided  of ghost-free effective theories
with higher dimensional (derivative) {\it interactions}:
a chiral superfield model with such an interaction operator, 
a model generating at low-energy  Dirac-gaugino masses and 
an effective  model  that is a  supersymmetric version of the  
Euler-Heisenberg Lagrangian.
Each model was  ``unfolded'' into a second-order theory for which we subsequently
 identified a 
 UV formulation ({\it i.e.} with two-derivatives only and ghost-free).

We also showed how our method can be extended to any $K$ and $W$
as arbitrary functions of chiral (functions) arguments. 
Finally, the Appendix provided  a special case showing  how
our method can be used  to all orders  in momentum expansion. 
This is relevant for momenta closer to the effective 
cutoff. This was possible  by using an iteration procedure and an infinite
set of superfield constraints, with similar conclusions as in the
examples quadratic in fields.

\vspace{2cm}
\setcounter{equation}{0}
\def\theequation{A-\arabic{equation}}
\setcounter{equation}{0}
\def\theequation{A-\arabic{equation}}
\def\thesection{A}
\setcounter{section}{0}

\section{Appendix: ``Unfolding'' effective operators to all orders}
\label{s6}
\label{ab}

In this Appendix we show that the ``unfolding'' method of using superfield constraints to
eliminate higher powers of the  superderivatives can be extended to all orders
in the momentum  expansion of an effective theory.

The plan of this section is as follows.
In Section~\ref{a1}, we choose a simple  model for which the UV completion is 
known, integrate a massive state $\chi$ {\it exactly} (to all orders),
 to generate an infinite series
in momentum expansion and a non-local theory, 
see eqs.(\ref{hhh}), (\ref{tt}), (\ref{tw}).
In section~\ref{a2} this theory truncated to an arbitrary order ($n$)
 is ``unfolded'' into a traditional second-order
theory that is shown to have only polynomial terms in superfields
and an additional  set of massive superfields (half of which are ghost-like),
eqs.(\ref{fff}), (\ref{f44}). This theory is classically equivalent to the starting
one of eqs.(\ref{tt}), (\ref{tw}) in the corresponding order in $\partial/\Lambda$.
 When integrating these massive
superfields one can then generate an effective Lagrangian of second-order, 
with only polynomial  effective operators as done in the text, eqs.(\ref{eer}), (\ref{eer2}).

\def\thesubsection{A.1}
\subsection{Effective operators from a renormalizable theory: all orders analysis}
\label{a1}

Consider a simple renormalizable, UV complete  Lagrangian of eq.(\ref{i1}) 
\medskip
\bea
\cL=\int d^4\theta\Big\{ \Phi_0^\dagger\,\Phi_0 +\chi^\dagger \chi\Big\}
+
\Big\{\int d^2\theta \Big[ (1/8)\,\Lambda \chi^2 + (1/4)\,m\,\Phi_0 \chi +W(\Phi_0)\Big]+h.c.
\Big\}
\label{hhh}
\eea

\medskip\noindent
so $\Lambda$ is here the mass of $\chi$.
Ignoring contributions from $W$ (if any),
the masses of the scalar components of $\chi$, $\Phi_0$ are
 $(1/8)\,\Lambda(1\mp [1+4 m^2/\Lambda^2)^{1/2}]$.
For  $\Lambda\gg m$ we can integrate  out $\chi$ via its eq of motion
\medskip
\bea
- \overline D^2 \chi^\dagger +\Lambda\chi +m\Phi_0=0
\eea

\medskip\noindent
which has a (iterative) solution
\medskip
\bea\label{ry}
\chi=\frac{-m}{\Lambda}\Phi_0 +\frac{-m}{\Lambda^2}\,\overline D^2\Phi_0^\dagger 
+\frac{-m}{\Lambda^3} \overline D^2 \,D^2 \Phi_0+
\frac{-m}{\Lambda^4} \overline D^2 \,D^2 \overline D^2\Phi^\dagger_0+
\cdots
\eea

\medskip\noindent
This solution is used back in $\cL$ to integrate $\chi$.
Due to technical difficulties,
one always truncates such expansion to a fixed (low) order. 
One finds for example
\medskip
\bea\label{t0}
\!\cL_{\rm eff}\!\!&=&\!\!\!\! \int d^4\theta\,\,
\Phi_0^\dagger\,\big[1-\xi^2 \Box_*\big]\,\Phi_0
\\
\!&+&\!\!\!\!\!
\int d^2\theta \,
\Big[
\frac{-m\,\xi}{8}
\big((1+\xi^2)^{-1/2}\Phi_0\big)^2
+\frac{m\xi}{8}\,\Phi_0\Box_*\Phi_0
+W\big[\Phi_0/(1+\xi^2)^{1/2}\big]\Big]+{\rm h.c.}
\!+\!\cO\Big(\frac{1}{\Lambda^5}\Big)
\nonumber
\eea

\medskip\noindent
after a rescaling $\Phi_0\ra \Phi_0/(1+\xi^2)^{1/2}$ was made and where
\medskip
\bea
\xi=\frac{m}{\Lambda},\qquad \Box_*=\frac{\Box}{(\Lambda/4)^2},
 \qquad \Box\Phi_0=-16 \,\overline D^2 D^2\,\Phi_0.
\eea

\medskip\noindent
Note the presence of the $\Box$ operators in the F- and D-terms
investigated in the text, also leading to the presence of the 
extra ghost superfields. 

But we can go beyond the approximation of truncated series.
One finds the exact result
\smallskip
\bea\label{tt}
\!\!\cL\!\!&=&\!\!\!\!\int\!\!  d^4\theta\,
\Phi_0^\dagger\,\Big[1+\frac{m^2}{\Lambda^2}\, \Big(
1 + 
\sum_{n\geq 1}\frac{1}{\Lambda^{2 n}}\big[\overline D^2 D^2\big]^n
\Big)\Big]\,\Phi_0
\\
&+&\!\!
\int\!\! d^2\theta \,
\Big[\,\frac{-m^2}{8 \,\Lambda}\,\Phi_0\,
\Big(1+
\sum_{n\geq 1}
 \frac{1}{\Lambda^{2n}}\big[\overline D^2 D^2\big]^n
\Big) \,
\Phi_0 + W(\Phi_0)
\Big]\!+\!{\rm h.c.}
\nonumber
\eea

\medskip\noindent
or
\medskip
\bea\label{tq}
\!\!\cL\!\!\!\!&=&\!\!\!\!\!\!\!\int\!\!  d^4\theta\,
\Phi_0^\dagger\,
\Big[1\! +  \xi^2\frac{1-(-\Box_*)^n}{1+\Box_*}\Big]\,
\Phi_0
\!+\!
\Big\{\!
\int\!\! d^2\theta
 \Big[\frac{-\Lambda}{8}\xi^2\,\Phi_0\,\frac{1-(-\Box_*)^n}{1+ \Box_*}\,\Phi_0 
\!+\! W(\Phi_0)\Big]\!\!+\!{\rm h.c.}
\!\!\Big\} \quad
\eea

\medskip\noindent
where $n\ra \infty$; in this limit there is no ghost,
but  truncating $\cL$ to $n$ finite generates ghosts (artefacts).
One  ignores the terms  $\Box_*^n$
since integrating out $\chi$ of mass of order $\Lambda$
we effectively integrate momenta that are below this mass scale. 
Further, the F-term can be replaced, up
to a total space-time integral as follows:
$\Phi_0\,(1+\Box_*)^{-1}\Phi_0\ra 
[(1+\Box_*)^{-1/2}\,\Phi_0]^2$.
To see this expand $(\Box_*+1)^{-1}$ and  use repeated
integration by parts. One then rescales $\Phi\ra Z^{1/2}\Phi$, $\Phi^\dagger\ra \Phi^\dagger Z^{1/2}$
where $Z^{-1}=1+\xi^2\,(\Box_*+1)^{-1}$ to find
\medskip
\bea\label{tw}
\cL=\int  d^4\theta\,
\Phi_0^\dagger\,\Phi_0
+
\Big\{
\int\!\! d^2\theta \, \Big[-\frac{1}{8}\, \Lambda\,\big[(1-Z)^{1/2}\Phi_0 \big]^2
+ W(\sqrt Z\Phi_0)\Big]\!+\!{\rm h.c.}
\Big\} 
\eea
where
\bea
1-Z=\frac{\xi^2}{1+\xi^2+\Box_*}.
\eea

\medskip\noindent
Eq.(\ref{tw}) 
is an exact result of integrating out the massive field $\chi$ to {\it all orders} in 
$\partial/\Lambda$. The global effect is a wavefunction renormalization,
albeit in operatorial sense.
The presence of $\Box$  inside $W$ generates (ghost-free) derivative 
interactions. The derivative acting in the (denominator of the)
bilinear F-term  brings  ghost superfields (when Taylor expanded and truncated,
see Section~\ref{s2}). 
Its structure is {\it non-local} and follows from the sum of the whole  series,
as expected from the discussion in Introduction.
Note that this term is  equivalent to $(-\Lambda/8)\,\Phi_0 \,(1-Z)\,\Phi_0$.
In the familiar approximation   $\Box_*\ll 1$ one finds $\cL_{\rm eff}$ of
 eq.(\ref{t0}).

\medskip
\def\thesubsection{A.2}
\subsection{``Unfolding'' the effective operators to all orders}
\label{a2}

For further insight into the momentum expansion of an effective theory,
let us examine the result of  eq.(\ref{tw}) by using the  ``unfolding'' procedure. 
We would like to see if such ``unfolded'' version still exists 
to {\it all orders} in momentum, i.e. we do not
``truncate'' the  Lagrangian after integrating $\chi$ in (\ref{hhh}).
To this end, we use  as many  constraints as needed to enforce 
the solution in eq.(\ref{ry}).
Introduce
\medskip
\bea\label{qq}
\overline D^2\Phi_0^\dagger &=& m\,\Phi_1 \nonumber\\
\overline D^2\Phi_1^\dagger &=& m\,\Phi_2 \nonumber\\
\cdots & & \cdots\nonumber\\
\overline D^2\Phi_{n-1}^\dagger &=& m\,\Phi_{n}, \qquad {\rm etc.} 
\eea

\medskip\noindent
therefore\footnote{We used in (\ref{qq}) the scale $m$ for dimensional reasons.
Using instead the scale $\Lambda$ does not work since then the eigenvectors
of the matrix of kinetic terms 
 would not  be normalizable to unity (unitarity violation).}
\bea
\chi=-\xi \sum_{j\geq 0} \xi^j\,\Phi_j,\qquad {\rm where} \qquad \xi\equiv m/\Lambda.
\eea

\medskip\noindent
Define new Lagrange multipliers  superfields $\Sigma_i$ i=1,2...., 
that enforce constraints (\ref{qq}), so we have an equal number of $\Sigma_i$ and
$\Phi_i$ (with $i\not=0$).
Then integrating $\chi$ to  all
orders in momentum gives the Lagrangian below
\bigskip
\bea
\cL&=&\int d^4\theta \,\,\Big[\Phi_0^\dagger\,\Phi_0+
\xi^2\, \Big\vert\sum_{j\geq 0}^n \xi^j \Phi_j\Big\vert^2\,\Big]
+\Big\{
\int d^2\theta
\,\,\frac{1}{8}\,\Lambda\,\xi^2\,\Big(\sum_{j\geq 0}^n\xi^j  \Phi_j\Big)^2
+{\rm h.c.}\Big\}
\nonumber\\
&+&
\!\!\!\Big\{
\int d^2\theta \Big[\,\,\frac{-m}{4}\xi\Phi_0\,
\Big(\sum_{j\geq 0}^n\xi^j \Phi_j\Big)
-
\frac{\xi}{4}\, \sum_{j\geq 1}^{n} 
\big(\overline D^2\Phi_{j-1}^\dagger -m\,\Phi_{j}\big)\,\Sigma_{j}
+W(\Phi_0)\Big]
+{\rm h.c.} \Big\}\qquad
\eea

\medskip\noindent
or equivalently\footnote{
Instead of multiplying the constraint by $\xi$
one can alternatively demand $\Sigma\!\ra\! 0$ fast enough at large $\Lambda$.} 
\medskip
\bea
\label{fff}
\qquad
\cL&=&\int d^4\theta 
\Big[
\Phi_0^\dagger\Phi_0
+\xi^2\Big\vert\sum_{j\geq 0}^n\xi^j \Phi_j\Big\vert^2
+\xi \sum_{j\geq 1 }^{n} \Big(\Phi_{j-1}^\dagger \Sigma_{j}\!\!+{\rm h.c.}\Big)
\Big]
\\
&+&
\!\!\!
\int d^2\theta\, \Big\{\,
\frac{m\xi}{8}\,\Big[
- \Phi_0^2\,
+
2\sum_{j=1}^n \Phi_j\Sigma_j
+\Big(\sum_{j=1}^n\xi^j\Phi_j\Big)^2\Big]
+W(\Phi_0)\Big\}+{\rm h.c.}\qquad\qquad\qquad
\nonumber
\eea

\medskip\noindent
This is an ``unfolding'' of  Lagrangian (\ref{tq}) or (\ref{tw})
 to all orders in ($\partial/\Lambda$) giving the result of
integrating out (exactly) the massive state 
$\chi$. Lagrangian (\ref{fff}) contains only renormalizable operators
by power  counting but also an infinite number of  superfields\footnote{This 
is not too surprising. The existence of higher powers of momenta in the 
expansion (i.e. more derivatives) demands more initial conditions (parameters), 
in this case extra fields. }  ($n\ra \infty$).
Truncating the number of superfields to finite $n$  is equivalent to a 
truncation of the momentum expansion of the solution $\chi$ 
and of the Lagrangian in eq.(\ref{tt}).
Note the polynomial form of  $\cL$.

Let us diagonalize the hermitian form of the 
kinetic terms  in the first line  of $\cL$ in eq.(\ref{fff}),
in the basis $(\Phi_0,\Phi_1,....,\Phi_n, \Sigma_1, \Sigma_2, ..., \Sigma_n)$,
$n\ra \infty$.
We  obtain a squared $(2n+1)\!\times\! (2n+1)$ matrix  $A_n$  ($n\ra\infty$). 
One can show that this matrix  has 
$\det A_n=(-1)^n\, \xi^{4 n+2}$.
Since  $\det A_n$ changes sign under $n\ra n+1$, 
each level $n$ (i.e. new constraint) adds an extra  ghost and 
an extra particle superfields. The characteristic equation is
\bea
\det (A_n-\lambda I_n)=-(\lambda-\xi)^{n-2} (\lambda+\xi)^{n-2}\,\cP(\lambda,n)=0
\eea
where  $n\geq 1$ and
\bea
\cP(\lambda, n)=\lambda^5
+
c_4\lambda^4+
c_3\lambda^3+
c_2\lambda^2+
c_1\lambda+
c_0
\eea
with
\bea
c_4&=&-\frac{1-\xi^{2n+4}}{1-\xi^2},\quad
c_3=\xi^2\,\Big(\frac{1-\xi^{2n +2}}{1-\xi^2}-3\Big),\quad
c_2= \xi^2\,\Big(\frac{1-\xi^{2n+4}}{1-\xi^2}+\xi^{2n+2}\Big),\,\,
\nonumber\\[8pt]
c_1&=&\xi^4\,(1-\xi^{2n}),\quad
c_0=-\xi^{2n+6}
\nonumber
\eea

\medskip\noindent
 The roots of $\cP(\lambda, n)=0$ are
\bea
\lambda_0&=& 1+2\,\xi^2 -2\,\xi^4+\cO(\xi^5),\qquad\,\,\,
\lambda_1=\xi^{2n+2}+\cO(\xi^{2n+3});\,\,\,\,\,
\nonumber\\[2pt]
\lambda_{2}&=&-\xi^2+2\,\xi^4+\cO(\xi^5),\qquad\quad\,\,\,
\lambda_{3,4}=\pm \xi+(1/2)\,\xi^4+\cO(\xi^5)
\eea

\medskip\noindent
Of the roots $\lambda_{0,1,2,3,4}$,  we see that 2 
are ghost-like and 3 (or 2) are particle-like for $n$ finite (infinite)
respectively. In addition we have pairs  of ghost
 and particle-like superfields ($\lambda=\pm\xi$) each of 
degeneracy $n-2$, as obvious from the characteristic equation.

Also note $\cP(\lambda,1)$ contains a factor
$(\lambda^2-\xi^2)$, so 
$\det(A_1-\lambda I_1)=-[\lambda^3-\lambda^2 (1+\xi^2+\xi^4)
-\lambda\,\xi^2\,(1-\xi^2)+\xi^6]$. The roots are in this case $\lambda_{0,1,2}$ shown 
above (with $n=1$), so we have 2 particle and 1 ghost-like superfields.
When truncating $\cL$ of eq.(\ref{fff}) to $n=1$
and after eliminating $\Sigma_1$ via its eqs of motion, we obtain exactly 
eq.(\ref{t0}), as expected. Thus eq.(\ref{fff}) for $n=1$ is a ``polynomial''
 version of eq.(\ref{t0}). This case was studied in Section~\ref{s2} 
and  lead to final  eqs.(\ref{eer}), (\ref{eer2}) after integrating the 
(massive)  ghosts.

The eigenvector  corresponding to the 
eigenvalues $\lambda_k$, $k$ {\it fixed} to  $k=0,1,2,3,4$, is
\medskip
\bea
(u^\dagger)_{zk}&\equiv&
N_k\,\big(1,\, \sigma_k\,\xi,\, ...,\sigma_k\,\xi^{n-1},\, \sigma_k^\prime\,\xi^n;\, \,
\,\xi/\lambda_k,\,\,
\sigma_k\,\xi^2/\lambda_k,\, ...,\,\sigma_k\,\xi^n/\lambda_k\big)^T,
\\[7pt]
\sigma_k&\equiv& \frac{\lambda_k^2-\xi^2-\lambda_k}{\lambda_k^2-\xi^2},\quad
\sigma_k^\prime=\frac{\lambda_k^2-\lambda_0-\xi^2}{\lambda^2_k}, \qquad
k=0,1,2,3,4; \,\,\,\, z=0,1,2,\cdots,2n.
\nonumber
\eea

\medskip\noindent
$N_k$ is a constant of normalization
to unity of $(U^\dagger)_{jk}$ where $k$ is fixed.
With $\vert\xi\vert<1$  ($\vert m\vert<\Lambda$), these
eigenvectors ($k$ fixed)
are indeed normalizable  for $n\ra \infty$ ($N_k<\infty$),
so unitary is not violated. 
Further,  for an eigenvalue  $\lambda=\xi$, for finite $n$ the eigenvector is
$(0, \Phi_1, ...,\Phi_{n-1}, 0, 0, \Phi_1,..,\Phi_{n-1})^T$  
and if $\lambda=-\xi$, it is 
$(0, \Phi_1,..,\Phi_{n-1}, 0, 0, -\Phi_1,...,-\Phi_{n-1})^T$.
In both cases $\lambda=\pm\xi$, the fields $\Phi_j$ are arbitrary
up to the constraint  $\sum_{j=1}^{n-1}\xi^j\Phi_j=0$; one can 
choose any two fields to implement this constraint
with the remaining fields set to 0.

The Lagrangian in the diagonal basis becomes, after a rescaling
$\tilde\Phi_z\ra \tilde\Phi_z/\sqrt{\vert\lambda_z\vert}$: 
\bea
\label{f44}
\cL&=&\int d^4\theta 
\,\Big[\,\,\tilde \Phi_0^{\dagger}\,\tilde\Phi_0
\,+
\,\tilde\Phi_{j}^{\dagger}\tilde\Phi_{j}\,-\,\tilde\Phi_{j+n}^\dagger\tilde\Phi_{j+n}\Big]
+\!\!\! \int d^2\theta\,\, 
\Big[\, 
U_{zz^\prime}\frac{
\tilde\Phi_z\tilde\Phi_{z^\prime}}{\sqrt{\vert \lambda_z\lambda_{z^\prime}\vert}}
+
W\Big(\frac{u^\dagger_{0z}\tilde\Phi_z}{\sqrt{\vert\lambda_z\vert}}\Big)\Big]+{\rm h.c.}
\nonumber
\eea

\medskip\noindent
where
\bea
U_{zz^\prime}= \frac{m\xi}{8}\,
\big[- u^\dagger_{0\,z} u^\dagger_{0 z^\prime}\!\!+ 2\,
u^\dagger_{jz} u^\dagger_{j+n, z^\prime}
+
\xi^j \,\xi^{k}\, u^\dagger_{jz}\, u^\dagger_{k z^\prime}
\big]
\eea
and where sums (not shown) are understood over the repeated indices ($n$ fixed),
with
\medskip
\bea
j, k=1, 2,...., n; \,\,\,
z, z^\prime=0, 1, 2,..., 2 n;\quad
\lambda_z=\{\,\lambda_0, \lambda_1,\,\lambda_3,\,\underbrace{\,\xi,....,\xi\,}_{n-2}; \,
\lambda_2,\,\lambda_4,\,\underbrace{-\xi,....,-\xi\,}_{n-2}\,\}.
\eea
$\tilde\Phi_{n+1},...,\tilde\Phi_{2n}$ are $n$ the ghost superfields.
From the F-term bilinears and ignoring contributions from $W$,
 it can be shown that one state is  light (original particle $\Phi_0$)
while the other (particle and ghost-like) superfields
are massive (mass of order $\Lambda$).

All superderivatives generated after the integration of 
massive superfield $\chi$ were eliminated.
The above Lagrangian  has only interactions polynomial
in superfields and all  operators of $d>4$ were eliminated via superfield
constraints\footnote{A similar example exists  for the 
effective Akulov-Volkov action for the goldstino; this action can be completely
 expressed in terms of superfields, in a interaction-free theory 
$L=\int d^4\theta G^\dagger G+\int d^2\theta f\, G+h.c.$,  $f\not=0$, 
endowed with the constraint $G^2=0$ where $G$ is the 
goldstino superfield \cite{SK,SK1,SK2,SK3,brignole,SK4}.}.
The downside is the presence of an infinite set of fields, all massive, 
beyond initial $\Phi_0$.
These can be integrated out as we did in the ``truncated'' case.
This description is classically  equivalent to $\cL$ of eq.(\ref{hhh}), (\ref{tw})
and may be useful in applications. 
This method can also be applied to more complicated cases.

\vspace{2cm}
\noindent
{\bf Acknowledgements:  }
E. Dudas thanks the Alexander von Humboldt foundation 
and DESY-Hamburg for support and hospitality in the final stages of this work.
The work of D.~Ghilencea was supported by the  Romanian Research Council
under  project number PN-II-ID-PCE-2011-3-0607 of the
National  Authority for Scientific Research, CNCS-UEFISCDI.

\end{document}